\documentclass[acmtog]{acmart}

\setcopyright{cc}
\setcctype{by-nd}
\acmJournal{TOG}
\acmYear{2025} \acmVolume{44} \acmNumber{6} \acmArticle{222}
\acmMonth{12} \acmPrice{}\acmDOI{10.1145/3763328}

\usepackage{lineno}
\usepackage{booktabs}
\usepackage{graphicx}
\usepackage{microtype}
\usepackage{collect}
\usepackage{xspace}
\usepackage{wrapfig}
\usepackage{bbm}
\usepackage{soul}

\newcommand{\refFig}[1]{Fig.~\ref{fig:#1}}
\newcommand{\refTab}[1]{Tab.~\ref{tab:#1}}
\newcommand{\refSec}[1]{Sec.~\ref{sec:#1}}
\newcommand{\refEq}[1]{Eq.~\ref{eq:#1}}

\newcommand{\dataset}[1]{\textsc{#1}}

\definecollection{mymaths}

\newcommand{\mymath}[2]{
    \newcommand{#1}{\TextOrMath{$#2$\xspace}{#2}}
    \begin{collect}{mymaths}{}{}{}{}
    #1
    \end{collect}
}

\mymath{\implicitfct}{f}
\mymath{\shape}{\mathcal{S}}
\mymath{\coord}{\mathbf{x}}
\mymath{\R}{\mathbb{R}}
\mymath{\Rtwo}{\R^2}
\mymath{\Rthree}{\R^3}
\mymath{\trainableparams}{\theta}
\mymath{\neuralimplicitfct}{\implicitfct_\trainableparams}
\mymath{\numlayers}{L}
\mymath{\layerindex}{l}
\mymath{\numneurons}{d}
\mymath{\weights}{W}
\mymath{\bias}{\mathbf{b}}
\mymath{\preneuron}{\mathbf{p}}
\mymath{\postneuron}{\mathbf{q}}
\mymath{\activation}{\sigma}
\mymath{\neuronindex}{i}
\mymath{\coordencoding}{\gamma}
\mymath{\freqvec}{\mathbf{a}}
\mymath{\numfreqs}{n}
\mymath{\domain}{\Omega}
\mymath{\cell}{\mathcal{C}}
\mymath{\cellvertices}{\widetilde{V}}
\mymath{\numcellvertices}{m}
\mymath{\cellweights}{\widetilde{\weights}}
\mymath{\cellbias}{\widetilde{\bias}}
\mymath{\cellfct}{\widetilde{\preneuron}}
\mymath{\cellvertex}{\mathbf{v}}
\mymath{\mask}{\mathbf{m}}
\mymath{\chordvertex}{\mathbf{c}}
\mymath{\chordX}{x}
\mymath{\chordY}{y}
\mymath{\activationapprox}{\activation_\chordvertex}
\mymath{\numchords}{M}
\mymath{\chordsegmentslope}{\alpha}
\mymath{\chordsegmentintercept}{\beta}

\newcommand{\eg}{e.g.\ }
\newcommand{\ie}{i.e.\ }

\newcommand{\baseline}[1]{{#1}}
\newcommand{\neurarch}[1]{{#1}}
\newcommand{\metric}[1]{{#1}}

\definecolor{MCColor}{rgb}{0.556,0.42,0.9}
\definecolor{DCColor}{rgb}{0.12,0.45,0.76}
\definecolor{HMCColor}{rgb}{0.274,0.74,0.77}
\definecolor{ArcColor}{rgb}{0.26,0.78,0.19}
\definecolor{AM}{rgb}{1,0.6,0}
\definecolor{ES}{rgb}{1,0,0.6}
\definecolor{Ours}{rgb}{1,0,0}

\newcommand{\mcbullet}{\textcolor{MCColor}{\Large \textbullet}}
\newcommand{\dcbullet}{\textcolor{DCColor}{\Large \textbullet}}
\newcommand{\hmcbullet}{\textcolor{HMCColor}{\Large \textbullet}}
\newcommand{\arcbullet}{\textcolor{ArcColor}{\Large \textbullet}}
\newcommand{\ambullet}{\textcolor{AM}{\Large \textbullet}}
\newcommand{\esbullet}{\textcolor{ES}{\Large \textbullet}}
\newcommand{\oursbullet}{\textcolor{Ours}{\Large \textbullet}}

\title[Marching Neurons]{Marching Neurons:\\Accurate Surface Extraction for Neural Implicit Shapes}

\author{Christian Stippel}
\orcid{0000-0003-0482-902X}
\affiliation{%
  \institution{TU Wien}
  \country{Austria}
}

\author{Felix Mujkanovic}
\orcid{0009-0009-9122-4408}
\affiliation{%
  \institution{Max-Planck-Institute for Informatics}
  \country{Germany}
}

\author{Thomas Leimkühler}
\orcid{0009-0006-7784-7957}
\affiliation{%
  \institution{Max-Planck-Institute for Informatics}
  \country{Germany}
}

\author{Pedro Hermosilla}
\orcid{0000-0003-3586-4741}
\affiliation{%
  \institution{TU Wien}
  \country{Austria}
}

\begin{document}

\begin{abstract}
Accurate surface geometry representation is crucial in 3D visual computing. Explicit representations, such as polygonal meshes, and implicit representations, like signed distance functions, each have distinct advantages, making efficient conversions between them increasingly important. Conventional surface extraction methods for implicit representations, such as the widely used Marching Cubes algorithm, rely on spatial decomposition and sampling, leading to inaccuracies due to fixed and limited resolution.  
We introduce a novel approach for analytically extracting surfaces from neural implicit functions. Our method operates natively in parallel and can navigate large neural architectures. By leveraging the fact that each neuron partitions the domain, we develop a depth-first traversal strategy to efficiently track the encoded surface. The resulting meshes faithfully capture the full geometric information from the network without ad-hoc spatial discretization, achieving unprecedented accuracy across diverse shapes and network architectures while maintaining competitive speed.
\end{abstract}

\begin{teaserfigure}
  \includegraphics[width=\linewidth]{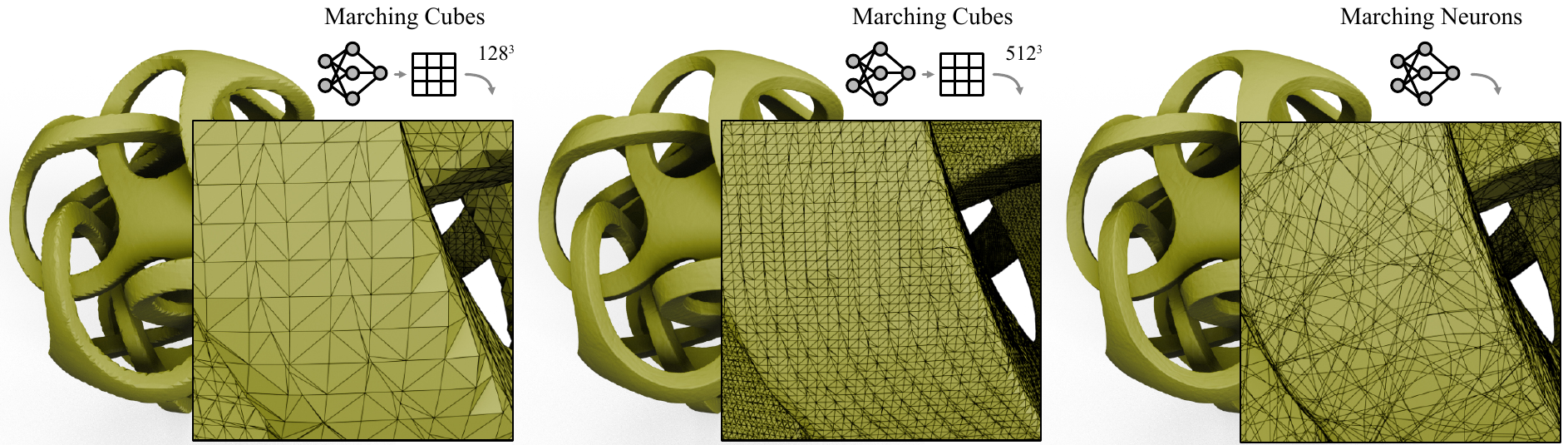}
  \Description{Teaser figure showing the mesh reconstruction quality of our method compared to sample-based methods such as Marching Cubes.}
  \caption{Surfaces extracted from a signed distance function (SDF) represented by a neural network using Marching Cubes with different grid resolutions (left and center) compared to our analytic method (right).
  While Marching Cubes struggles to reconstruct sharp edges even at high grid resolutions, our analytic method is able to reconstruct the surface accurately.}
  \label{fig:teaser}
\end{teaserfigure}

\maketitle


\section{Introduction}

Accurately representing surface geometry is a cornerstone of 3D visual computing and beyond~\cite{watt19993d,de2000computational}.
A continually growing repertoire of representations exists, broadly divided into two categories: 
\emph{Explicit} representations, which directly describe the surface using geometric elements such as polygons or points; 
and \emph{implicit} representations, which define the surface indirectly as the set of points satisfying a mathematical equation, such as level sets, with signed distance functions (SDFs) being a prominent example.
The choice of representation is typically guided by the downstream algorithms and applications that use the geometry.
For example, explicit representations are well-suited for classical tasks such as rendering~\cite{akenine2019real} and editing~\cite{botsch2010polygon}, while implicit representations excel in gradient-based geometry optimization, a core operation in modern neural workflows~\cite{park2019deepsdf,wang2021neus}.
In this work, we propose a novel, accurate, and flexible technique for analytically converting a neural implicit shape into an explicit surface.

The vast majority of existing techniques for surface extraction from implicit representations rely on spatial decomposition and sampling~\cite{de2015survey}.
The most commonly used method in this domain is Marching Cubes~\cite{lorensen1987marching}, which samples the implicit function on a regular grid and uses linear interpolation to approximate the desired level set.
While popular for its simplicity and agnostic to the implicit function representation, this method's accuracy is limited by the chosen sampling resolution and interpolation scheme.
Many follow-up works have further developed the basic approach~\cite{NEWMAN2006854}, \eg by pruning samples using hierarchical structures~\cite{wilhelms1992octrees}, moving to tetrahedral meshes to enable irregular sample placement~\cite{doi1991efficient}, optimizing the sample locations~\cite{shen2023flexicubes}, or employing more advanced interpolation strategies~\cite{ju2002dual,Sellan2024RFTA}.
Yet, for all of these approaches, a fundamental problem remains: 
Reconstructing complex geometry from a finite set of samples inevitably introduces inaccuracies.

The surge of deep learning~\cite{Goodfellow-et-al-2016} over the past decade has advanced not only signal processing but also signal representation. In particular, neural fields~\cite{xie2022neural,sitzmann2019scene}, which use coordinate-based networks for signal representation, have become a powerful and ubiquitous paradigm for continuous functions, such as surface implicits~\cite{park2019deepsdf}. The key insight of this work is that these neural representations enable \emph{efficient analytic surface extraction} without the need for ad-hoc sampling.

We propose a novel method for extracting a surface from an implicit function represented by a neural network.
We leverage the fact that a (deep) composition of piecewise linear functions remains piecewise linear, with each layer subdividing existing regions~\cite{montufar2014number} and shaping the target function through convex linear regions separated by hyperplanes.
This enables us to traverse the deep network neuron by neuron (``marching'') in a depth-first fashion, identifying progressively finer regions that contain the desired surface.
Although the number of linear regions increases exponentially with network depth~\cite{montufar2014number}, most do not contain the surface. 
Using range analysis~\cite{sharp2022spelunking} to eliminate empty regions, our approach efficiently traverses ReLU-based neural architectures. 
This process produces a polygonal mesh that accurately captures the encoded surface geometry.

Unlike previous work on analytic surface extraction~\cite{Lei2020,Lei2021,Berzins23}, our approach is not only adaptively accurate -- extracting surfaces at the level of the underlying neural field -- but also capable of extracting multiple disconnected shapes while maintaining competitive reconstruction speed. 

In summary, our contributions are:
\begin{itemize}
    \item A novel method for analytically extracting a polygonal mesh from an implicit neural representation.
    \item An algorithm that is natively parallel and easy to implement.
    \item Capabilities and accuracy that significantly exceed the current state of the art.
\end{itemize}

\noindent
We provide all source code and datasets in our \href{https://phermosilla.github.io/neurons/}{project page}.


\section{Related Work}

\subsection{The Geometry of Neural Networks}

Deep neural networks are universal function approximators in theory~\cite{hornik1989multilayer} and demonstrate the ability to represent complex functions across diverse domains in practice~\cite{Goodfellow-et-al-2016}.
This remarkable flexibility arises from their compositional structure, which enables exponential expressivity with respect to the number of layers~\cite{montufar2014number}, allowing virtually any data topology to be handled~\cite{naitzat2020topology}.
Networks with ReLU activation functions are particularly well-suited for analyzing this property~\cite{pascanu2013number,raghu2017expressive,hanin2019deep}, as they model continuous piecewise linear functions. 
These networks partition the input domain into polyhedral regions, formed by an arrangement of folded hyperplanes that correspond to the decision boundaries of neurons~\cite{vallin2023geometric,grigsby2022transversality}.
We propose a novel algorithm for tracing the geometry of a neural network to extract an explicit representation from the implicit function it encodes.

Algorithms for analytically extracting the polyhedral complex of a ReLU network typically rely on mixed-integer linear programming~\cite{serra2018bounding}, neuron state flipping~\cite{Lei2020, Lei2021}, or recursive intersection and cutting~\cite{wang2022estimation,Humayun_2023_CVPR}, with applications in, \eg, safety verification~\cite{vincent2021RPM} and robustness~\cite{hein2019relu}.
However, the computational complexity of these approaches is typically substantial due to the combinatorial explosion in the number of polyhedra.
Addressing this problem, \citet{Berzins23} proposes increasing efficiency by eliminating redundancy through an edge-centric approach.
Our key insight is that, for the critical task of surface extraction from neural implicit representations, a massively parallel implementation can be achieved by a bespoke depth-first traversal of the network.
This approach results in accuracy that surpasses the state of the art, while maintaining competitive speed.


\subsection{Iso-Surface Extraction}

The representation of surfaces as level sets of implicit functions has a rich history across various scientific disciplines and decades~\cite{bloomenthal1997introduction, osher2004level}.
Extracting an explicit (typically polygonal) surface from this representation is a well-studied problem for which several classes of algorithms have been developed~\cite{de2015survey}.

While surface tracking~\cite{hilton1996marching} and shrink-wrapping approaches~\cite{stander1997guaranteeing, van2004shrinkwrap, hanocka2020point2mesh} have been applied with some success, the majority of research focuses on spatial decomposition techniques~\cite{bloomenthal1988polygonization}.
Here, space is divided into cells, and polygons are constructed within each cell containing the surface, typically by interpolating discrete samples.
The popular Marching Cubes algorithm~\cite{lorensen1987marching, wyvill1986data} and numerous follow-up works \cite{NEWMAN2006854, hege1997generalized, wilhelms1992octrees, montani1994discretized, chernyaev1995marching} use a regular grid of cubes.
The rigidity of this structure has been relaxed to allow for more adaptive discretizations, \eg by using tetrahedra~\cite{doi1991efficient, ren2025mcgrids}.
Unlike the sampling-based techniques, our method determines the geometry analytically.

Dual representations have been shown to outperform traditional approaches by offering more flexible and accurate surface reconstruction, especially in handling sharp features and complex topologies~\cite{ju2002dual,azernikov2005anisotropic,nielson2004dual}.
Furthermore, replacing hand-crafted extraction rules with learned ones~\cite{chen2021neural, chen2022neural} or adopting more sophisticated interpolation schemes~\cite{sellan2023reach, Sellan2024RFTA,kohlbrenner25} can lead to significant quality improvements.
By using an analytical approach, our algorithm eliminates the need for any sampling and interpolation schemes.

With the proliferation of neural fields~\cite{sitzmann2019scene,xie2022neural}, surface-encoding implicit functions are now commonly represented using neural networks~\cite{park2019deepsdf}.
Due to their closed-form nature, analytic surface extraction is feasible; however, the complexity of the underlying geometry presents significant challenges.
Analytic Marching~\cite{Lei2020, Lei2021} addresses this problem by explicitly enumerating the polyhedral cells that contain the surface, recursively visiting neighboring cells.
However, this approach faces challenges with multiple disconnected components.
\citet{Berzins23} addresses this problem by relying on an edge-centric approach instead of linear regions.
However, \citeauthor{Berzins23} struggles with complex architectures due to the lack of a filtering mechanism for empty regions, producing at the same time an excessive number of polygons.
In contrast, we propose a method capable of extracting shapes that can consist of multiple components while maintaining competitive speed.


\section{Background}
\label{sec:background}

In this section, we introduce the concepts and notation for neural implicit shape representations relevant to our approach. 

\begin{wrapfigure}{r}{0.16\textwidth}
    \vspace{-1mm}
    \begin{center}
        \includegraphics[width=0.16\textwidth]{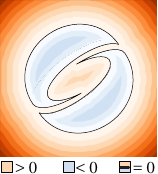}
    \end{center}
    \vspace{-2mm}
    \caption{An implicit function and its zero level set.}
    \label{fig:levelset}
\end{wrapfigure}

We focus on solid shapes \shape represented as the level set, or iso surface, of an implicit function
$\implicitfct \in \domain \rightarrow \R$ \cite{bloomenthal1997introduction}, with 
$ \domain \subset \Rthree$.
Without loss of generality, we consider the zero level set~(\refFig{levelset}):
\begin{equation}
    \shape := 
    \left\{
        \coord \in \domain \mid \implicitfct(\coord) = 0
    \right\}.
\end{equation}
This general setting encompasses several widely used special cases, including signed distance fields (SDFs)~\cite{osher2004level,park2019deepsdf}, unsigned distance fields~\cite{chibane2020neural}, and their variations, such as truncated~\cite{curless1996volumetric}, distorted~\cite{seyb2019non}, or otherwise weaker forms~\cite{sharp2022spelunking,marschner2023constructive}.

Nowadays, a widely adopted approach to representing \implicitfct is through a neural field \neuralimplicitfct, \ie a neural architecture with trainable parameters \trainableparams, optimized via gradient descent~\cite{xie2022neural}.
In the simplest case, \neuralimplicitfct is represented by a Multilayer Perceptron (MLP).
The number of neurons in each layer 
$\layerindex \in \left\{1, \dots, \numlayers \right\}$
is denoted as $\numneurons_\layerindex$.
Given a weight matrix 
$
\weights^{(\layerindex)}
\in 
\R^{\numneurons_\layerindex \times \numneurons_{\layerindex-1}}
$,
a bias vector
$\bias^{(\layerindex)} \in \R^{\numneurons_\layerindex}$, 
and a non-linear activation function 
$\activation^{(\layerindex)} \in \R \rightarrow \R$
for each layer, the network recursively applies
\begin{equation}
\label{eq:mlp_recursion}
    \preneuron^{(\layerindex)}
    = 
    \weights^{(\layerindex)} \postneuron^{(\layerindex-1)} + \bias^{(\layerindex)},
    \quad\quad
    \postneuron^{(\layerindex)} 
    =
    \activation^{(\layerindex)} \left( \preneuron^{(\layerindex)} \right),
\end{equation}
where \activation is applied element-wise.
We set 
$\postneuron^{(0)} := \coord$ and 
$\neuralimplicitfct(\coord) := \preneuron^{(\numlayers)}$ and refer to $\preneuron^{(\layerindex)}_\neuronindex$ and $\postneuron^{(\layerindex)}_\neuronindex$ as the $i$-th \emph{pre-activation neuron} and \emph{post-activation neuron}, respectively.

\begin{figure*}
    \includegraphics[width=0.99\linewidth]{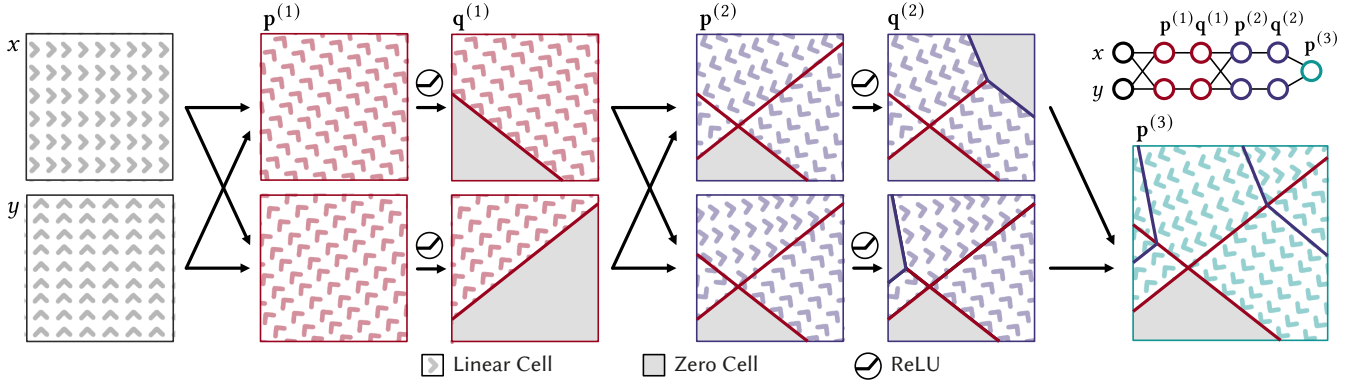}
    \caption{
    The geometry of ReLU MLPs, illustrated here for a 2D input domain with two hidden layers containing two neurons each. In each layer \layerindex, the pre-activations $\preneuron^{(\layerindex)}$ are computed as linear combinations of the neurons from the previous layer. The ReLU activation then produces post-activations $\postneuron^{(\layerindex)}$, introducing lines that subdivide the existing linear regions. As a result, the network's output is composed of convex cells within which it remains linear, \ie we can collapse all corresponding weight matrices and biases.}
    \label{fig:relu_geometry}
\end{figure*}

While there is an overwhelming variety of nonlinear activation functions available~\cite{kunc2024three}, we consider the special case of \emph{piecewise linear}\footnote{For simplicity, we use ``linear'' for both linear and affine functions.} activation functions for now.
Given the recursive structure of \refEq{mlp_recursion}, \neuralimplicitfct is a composition of piecewise linear functions and is therefore itself piecewise linear.
Each call of \activation potentially subdivides the domain of \neuralimplicitfct into an increasingly larger number of convex linear regions, separated by planes~\cite{montufar2014number,hanin2019deep}, with each region referred to as a cell.
\refFig{relu_geometry} illustrates this property in 2D using arguably the most common choice in this space: the rectified linear unit (ReLU):
\begin{equation}
\label{eq:relu}
    \activation ( \preneuron ) 
    =
    \max(0, \preneuron).
\end{equation}
In this context, we refer to a post-activation neuron $\postneuron^{(\layerindex)}_\neuronindex$ as \emph{active} if it is positive and \emph{inactive} if it is zero.
The cells of \neuralimplicitfct are separated by the set of planes
$
\left\{
    \coord \in \Rthree \mid \preneuron^{(\layerindex)}_\neuronindex(\coord) = 0
\right\},
$
which correspond to the locations where a neuron switches from inactive to active. 
Within each cell, the active/inactive pattern of neurons in the entire network remains fixed, implying that \neuralimplicitfct is linear within each cell and can be represented by collapsing the corresponding submatrices of $\weights^{(\layerindex)}$ and subvectors of $\bias^{(\layerindex)}$ across layers.
Similar observations hold for any choice of \activation that is piecewise linear.

\section{Method}

Given an implicit neural representation \neuralimplicitfct with ReLU activation functions, we aim to extract its zero level set \shape as a polygonal mesh. 
Importantly, our goal is an \emph{analytic} extraction that captures all the details encoded in \neuralimplicitfct, unlike the widely used sampling-based methods~\cite{lorensen1987marching}.

For piecewise-linear activation functions, \neuralimplicitfct can be viewed as a collection of linear branches~\cite{Lei2020,Lei2021}. 
Our key insight is that by adaptively subdividing the domain and tracking activation patterns, we can efficiently reduce the network to an explicit piecewise-linear function in each cell, represented as a polyhedral mesh. 
Using range analysis~\cite{sharp2022spelunking}, we can discard many cells early on, as they do not contain the zero level set, leading to significant efficiency gains. 
As a result, we obtain an accurate solution for the zero iso-surface in the form of a polygonal mesh.

In \refSec{2drelu}, we introduce the basic structure of our approach using a 2D domain, before extending the method to the 3D setting in \refSec{3dfull}.
Finally, we provide implementation details in \refSec{implementation}.


\subsection{Level Set Extraction for 2D Networks}
\label{sec:2drelu}

For $\domain \subset \Rtwo$, each neuron can define a polyline that partitions \domain into two regions, resulting in a collection of convex polygonal cells (\refFig{relu_geometry}).
We propose a novel scheme that explicitly tracks these cells by a depth-first traversal of \neuralimplicitfct.
As we traverse \neuralimplicitfct, the cells, with layer-specific geometries and internal states, undergo transformations and splitting, and may also be discarded.
The goal of this process is to identify the set of cells containing the zero level set of \neuralimplicitfct, enabling its explicit representation as one or more closed line strips to be extracted.

Each polygonal cell at layer \layerindex is represented as the tuple  
\begin{equation}  
    \cell^{(\layerindex)} = \left\{ \cellvertices; \cellweights, \cellbias \right\},  
\end{equation}  
where 
$
\cellvertices 
= 
\left[ \cellvertex_1, \cellvertex_2, \dots, \cellvertex_\numcellvertices \right] 
\in \R^{2 \times \numcellvertices}
$
specifies the 2D coordinates of the \numcellvertices vertices
$\cellvertex_j \in \Rtwo$
forming the polygon.
Further, $\cellweights \in \R^{\numneurons_\layerindex \times 2}$ and 
$\cellbias \in \R^{\numneurons_\layerindex}$  
are the parameters defining the linear function  
\begin{equation}
\label{eq:cell_fct}
    \cellfct(\coord) = \cellweights \coord + \cellbias  
\end{equation}  
inside the polygon by combining the linear functions contributing up to this layer, corresponding to the pre-activation response of the current layer.
For any hidden layer \layerindex, $\cellfct(\coord)$ produces a $\numneurons_\layerindex$-dimensional vector, where each component corresponds to a neuron (rows in \refFig{relu_geometry}).
In the final layer \numlayers, $\cellfct(\coord)$ outputs a scalar ($\numneurons_\numlayers = 1$) that represents the linear function describing the full \neuralimplicitfct within the cell.

Without loss of generality, we assume a rectangular domain \domain and initialize the cell set with a single quadrilateral, 
$\left\{ \cell^{(0)}_1 \right\}$, 
covering the entire domain. The corresponding function $\cellfct(\coord)$ is initialized as the identity function, with  
$\cellweights = \mathbb{I}_{2 \times 2}$ and $\cellbias = \mathbf{0}$.

\begin{figure}
    \includegraphics[width=0.99\linewidth]{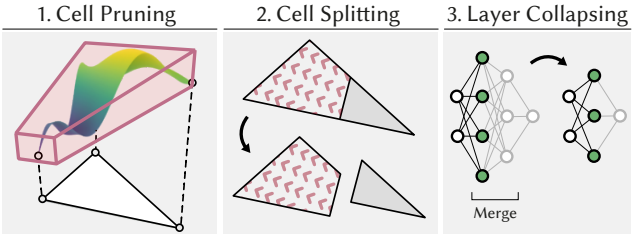}
    \caption{The three steps of our network traversal. First, we bound the network response within each cell (purple volume) to eliminate those that do not intersect the zero level set. Next, we split the remaining cell to capture the nonlinearities introduced by piecewise linear activation functions. Finally, the current layer (green neurons) is combined with the next layer. The entire process (1. - 3.) repeats until the final network layer is reached.}
    \label{fig:overview}
\end{figure}

From this point, we traverse the layers of \neuralimplicitfct, iteratively updating the cell set until only the final-layer cells containing the zero level set remain.
Progressing through layers, we prune or split cells while updating their parameters, as detailed below and shown in \refFig{overview}.

\subsubsection{Step 1: Cell Pruning}
\label{sec:pruning}
As a first step, we eliminate cells that do not intersect the zero level set of \neuralimplicitfct.  
Since the number of linear regions within the domain \domain grows exponentially with each layer~\cite{montufar2014number}, and most do not contain the zero level set, this step is crucial for maintaining efficiency.
We use range analysis~\cite{duff1992interval, rump2015implementation} on the remaining layers to accomplish this task.
Specifically, we follow \citet{sharp2022spelunking} and compute affine bounds~\cite{ld1993affine} of \neuralimplicitfct within each cell $\cell^{(\layerindex)}_i$.
This gives us a conservative estimate of the minimum and maximum values of \neuralimplicitfct within $\cell^{(\layerindex)}_i$.
If
\begin{equation}
\label{eq:bounding}
    \min_{\coord \in \cell^{(\layerindex)}_i} \neuralimplicitfct(\coord) > 0
    \quad
    \text{or}
    \quad
    \max_{\coord \in \cell^{(\layerindex)}_i} \neuralimplicitfct(\coord) < 0,
\end{equation}
we discard $\cell^{(\layerindex)}_i$ because it does not intersect the zero level set.
In practice, we query the affine bounds using the cell's axis-aligned bounding box.
This may overestimate ranges for degenerated cells, for example, ones that have a long, not axis-aligned diagonal; however, it suffices to ensure scalability to larger networks.

\subsubsection{Step 2: Cell Splitting}
\label{sec:splitting}
In this step, we first identify neurons in the current layer \layerindex that divide the polygon $\cell^{(\layerindex)}_i$ into two smaller linear regions. 
We refer to these neurons as \emph{critical} and observe that their pre-activation responses $\cellfct(\coord)$ change sign within the polygon, leading to the non-linear kink of the subsequent ReLU activation occurring inside the polygon.
Since we are dealing with linear functions, the vertices $\cellvertex_j$ naturally correspond to the locations of the extrema of $\cellfct(\coord)$ within the polygon.
Therefore, to check for criticality, it suffices to evaluate \cellfct at the vertices. Specifically, neuron $i$ is considered critical if 
\begin{equation}
\label{eq:criticalcheck}
    \min_j \left( \cellfct(\cellvertex_j) \right)_i < 0
    \quad
    \text{and}
    \quad
    \max_j \left( \cellfct(\cellvertex_j) \right)_i > 0.
\end{equation}
If one or more critical neurons are associated with the polygon, it is split along the (linear) zero level set of the corresponding output dimensions of $\cellfct(\coord)$ with the Sutherland-Hodgman~\shortcite{sutherland1974reentrant} algorithm.
New vertices are inserted on the intersection of polygon edges with the cutting plane, obtaining two sub-polygons separated by the neuron.
Polygons that do not need to be split are retained unchanged.

\subsubsection{Step 3: Layer Collapsing}
\label{sec:collapse}
As we move to the next layer $\layerindex + 1$, it is necessary to update \cellweights and \cellbias.
Recall that only neurons active within a cell contribute their linear functions to it.  
To track this, we define a binary mask  
$\mask \in \R^{\numneurons_\layerindex}$,  
where each entry indicates whether the corresponding neuron in the current layer \layerindex is active within the cell.  
Using this mask, we update the function parameters of the cell as follows:
\begin{equation}
\label{eq:relu_collapse}
\begin{alignedat}{2}
    &\cellweights
    && := 
    \weights^{(\layerindex+1)} \, \text{diag}(\mask) \, \cellweights 
    \\
    &\cellbias
    && := 
    \weights^{(\layerindex+1)} \left( \mask \odot \cellbias \right)
    +
    \bias^{(\layerindex+1)},
\end{alignedat}
\end{equation}
where $\text{diag}(\cdot)$ converts a vector into its diagonal matrix representation, and $\odot$ denotes the Hadamard product.
The updated cells form the new set  
$\left\{ \cell^{(\layerindex+1)}_i \right\}$, 
which will be processed by step 1 again.

\subsubsection{Network Traversal}
The three steps outlined above specify how cells should be updated when transitioning from one layer to the next, providing flexibility in choosing the global traversal scheme. 
A straightforward approach processes all cells in a layer before moving to the next, corresponding to a breadth-first traversal.
However, this approach has the disadvantage of requiring more memory than is typically available on a contemporary GPU.
In the worst case, every neuron splits all active cells. 
A breadth-first traversal must then hold the full frontier of all cells at once, resulting in space requirements of $O(2^\numneurons)$, where $\numneurons$ is the number of neurons. 
Therefore, we choose a depth-first traversal where we process cells in a last-in-first-out principle to reduce the memory burden. 
To fully utilize the GPU, we schedule the traversal to maintain a sufficient number of cells processed in parallel in each step.
In this case, we only store the current branches of the subdivision tree, resulting in $O(b\numneurons)$, where $b$ is the number of cells processed in parallel.

In practice, the average space complexity is lower: not every neuron splits every cell, and the likelihood of a split decreases the more often a cell has already been subdivided. 
Additionally, range-analysis pruning removes many cells before they would be split.

\subsubsection{Level Set Extraction}
Once all cells have reached the final layer, the set 
$\left\{ \cell^{(\numlayers)}_i \right\}$
contains the cells that cover the piecewise linear regions of the level set of \neuralimplicitfct. 
At this point, the linear level set
$
    \shape = 
    \left\{
        \coord \mid \cellfct(\coord) = 0
    \right\}
$
can be analytically extracted from each cell. 
This is straightforward as \cellfct is a linear function.
The union of the extracted per-cell level sets forms the desired explicit representation as a line strip.


\subsection{Extension to 3D}
\label{sec:3dfull}


Extending our approach to 3D, where  
$\coord \in \domain \subset \Rthree$,  
is straightforward: 
operations on polygons are adapted to handle \emph{3D polyhedra}.  
Specifically, we now maintain \numcellvertices 3D vertices,  
$
\cellvertices \in \R^{3 \times \numcellvertices},
$
and the per-cell function parameter  
$
\cellweights \in \R^{\numneurons_\layerindex \times 3}  
$
accounts for the additional dimension.
Critical neurons now split polyhedra along \emph{planes}.
Once the final layer \numlayers of \neuralimplicitfct is reached, we analytically extract the zero level set from each polyhedron, producing the polygonal mesh that explicitly represents the encoded surface \shape.
The mesh is then tessellated into triangles for compatibility with standard pipelines.


\subsection{Implementation Details}
\label{sec:implementation}

We realized our approach using a reasonably optimized JAX~\cite{bradbury2018jax} implementation that takes advantage of our natively parallel algorithm design.
All active cells reside in a shared buffer with indexing stacks for each operation.
To prevent numerical inaccuracies that can occasionally arise from our deep recursive subdivision scheme, we found a mixed-precision implementation to be essential.
Specifically, while cell pruning can be safely executed with 32-bit floating-point precision, we use 64-bit precision for cell splitting and layer collapsing.
However, this does not impose any restrictions on the bit depth of the input network \neuralimplicitfct.


\begin{table*}[hbt!]
\centering
\caption{\textbf{Mesh extraction.} Soft Precision (SP) and Soft Recall (SR) are multiplied by $10^6$, triangles divided by $10^3$. Runtime is measured in seconds.}
\label{tab:main-results}
\setlength{\tabcolsep}{9pt}
\begin{tabular}{lrrrrrrrr}
    \toprule
		& \multicolumn{4}{c}{\neurarch{d4\_w128}} & \multicolumn{4}{c}{\neurarch{d4\_w256}}\\
		\cmidrule(l{2pt}r{2pt}){2-5}
		\cmidrule(l{2pt}r{2pt}){6-9}
		& \multicolumn{1}{c}{SP} & 
        \multicolumn{1}{c}{SR} & 
        \multicolumn{1}{c}{Runtime} & 
        \multicolumn{1}{c}{Triangles} & 
        \multicolumn{1}{c}{SP} & 
        \multicolumn{1}{c}{SR} & 
        \multicolumn{1}{c}{Runtime} & 
        \multicolumn{1}{c}{Triangles} \\

    \midrule

        \hspace{-.25cm} \vspace{-.1cm} \emph{Approx. - $64^3$} \\
        \midrule

        \mcbullet   \baseline{Marching Cubes}        & 1682.28 & 10293.05 & 0.01 & 14.44 & 1802.69 & 9832.66 & 0.01 & 14.21 \\
        \dcbullet   \baseline{Dual Contouring}       & 1278.87 & 10044.56 & 0.20 & 14.45 & 1412.82 & 9530.62 & 0.20 & 14.23 \\
        \hmcbullet  \baseline{Hier. Marching Cubes}  & 1686.56 & 10352.45 & 4.77 & 14.42 & 1814.43 & 9982.32 & 5.72 & 14.18 \\
        \arcbullet  \baseline{Reach for the Arcs}    & 4072.33 & 3144.99 & 538.65 & 355.01 & 3480.32 & 2851.51 & 481.15 & 349.09 \\

        \hspace{-.25cm} \vspace{-.1cm} \emph{Approx. - $128^3$} \\
        \midrule

        \mcbullet   \baseline{Marching Cubes}        & 518.48 & 1373.29 & 0.04 & 62.18 & 551.65 & 1248.17 & 0.08 & 60.89 \\
        \dcbullet   \baseline{Dual Contouring}       & 425.45 & 1319.47 & 1.17 & 62.13 & 461.12 & 1207.23 & 1.22 & 60.90 \\
        \hmcbullet  \baseline{Hier. Marching Cubes}  & 543.82 & 1420.48 & 8.83 & 62.11 & 561.32 & 1269.78 & 11.69 & 60.76 \\
        \arcbullet  \baseline{Reach for the Arcs}    & 6302.02 & 2810.83 & 5726.60 & 310.95 & 4826.59 & 2406.64 & 4696.52 & 337.86 \\

        \hspace{-.25cm} \vspace{-.1cm} \emph{Approx. - $256^3$} \\
        \midrule

        \mcbullet   \baseline{Marching Cubes}        & 162.36 & 356.69 & 0.30 & 258.50 & 175.50 & 330.28 & 0.62 & 252.57 \\
        \dcbullet   \baseline{Dual Contouring}       & 126.58 & 327.90 & 6.92 & 258.38 & 134.16 & 301.52 & 7.07 & 252.47 \\
        \hmcbullet  \baseline{Hier. Marching Cubes}  & 221.47 & 436.62 & 11.35 & 259.51 & 196.49 & 363.15 & 16.81 & 252.52 \\
        \arcbullet  \baseline{Reach for the Arcs}    & -- & -- & -- & -- & -- & -- & -- & -- \\

        \hspace{-.25cm} \vspace{-.1cm} \emph{Approx. - $512^3$} \\
        \midrule

        \mcbullet   \baseline{Marching Cubes}        & 53.61 & 113.88 & 2.36 & 1048.80 & 60.33 & 101.94 & 4.91 & 1025.29 \\
        \dcbullet   \baseline{Dual Contouring}       & 41.75 & 101.98 & 48.56 & 1048.69 & 45.24 & 88.62 & 50.33 & 1025.15 \\
        \hmcbullet  \baseline{Hier. Marching Cubes}  & 150.22 & 235.37 & 15.05 & 1060.95 & 103.01 & 153.12 & 21.54 & 1028.08 \\
        \arcbullet  \baseline{Reach for the Arcs}    & -- & -- & -- & -- & -- & -- & -- & -- \\

        \hspace{-.25cm} \vspace{-.1cm} \emph{Exact} \\
        \midrule

        \ambullet   \baseline{Analytic Marching}     & \textbf{0.03} & 675.76 & 3.21 & 425.92 & \textbf{0.02} & 188.80 & 32.99 & 2306.06 \\
        \esbullet   \baseline{Edge Subdivision}      & \textbf{0.03} & 0.61 & 46.72 & 429.08 & \textbf{0.02} & 0.90 & 21147.39 & 2321.70 \\
        \oursbullet \baseline{Ours}                  & \textbf{0.03} & \textbf{0.07} & 11.84 & 429.08 & \textbf{0.02} & \textbf{0.03} & 169.40 & 2321.70 \\

    \bottomrule
\end{tabular}
\end{table*}

\section{Evaluation}

In this section, we describe the experiments conducted to evaluate our method.
We compare the quality of the meshes extracted by our algorithm with those reconstructed by other methods.
Moreover, we also measure the time required for the reconstruction of the resulting meshes and the average number of triangles generated.
Additionally, we also evaluate the quality of the meshes generated and the effect of mesh simplification algorithms on the reconstruction.
Lastly, we evaluate the effect of our filtering step and extend our approach to arbitrary activation functions.
We assume that all shapes are defined by the zero-level set of an SDF encoded by a neural network.

\subsection{Experimental Setup}

\paragraph{Metrics}

To measure the error introduced by the reconstruction algorithms, we compare the extracted mesh directly to the SDF encoded within the neural network.
We define two metrics to measure such error: \metric{Soft-Precision} (SP) and \metric{Soft-Recall} (SR).
\metric{Soft-Precision} aims to quantify how far the reconstructed mesh is from the zero-level set defined by the SDF.
This metric is computed by sampling $2^{20}$ points on the surface of the reconstructed mesh, evaluating the SDF at these point locations, and taking the mean absolute value.
If the reconstructed surface lies in the zero-level set of the SDF, \metric{Soft-Precision} will be exactly zero.

However, \metric{Soft-Precision} does not capture cases in which large portions of the zero-level set are not reconstructed.
To remedy this, we introduce a second metric, \metric{Soft-Recall}.
First, we sample $2^{20}$ points on the surface of the original mesh from our dataset.
Then, we use gradient descent to move such points to the zero-level set of the SDF.
Once converged, we measure the average distance of the resulting coordinates to the reconstructed mesh.
If the method is able to perfectly reconstruct the complete zero-level set, \metric{Soft-Recall} will be exactly zero.

Additionally, we compare the time required to perform the extraction and the number of resulting triangles.

\paragraph{Datasets}

\begin{figure}
	\includegraphics[width=0.99\linewidth]{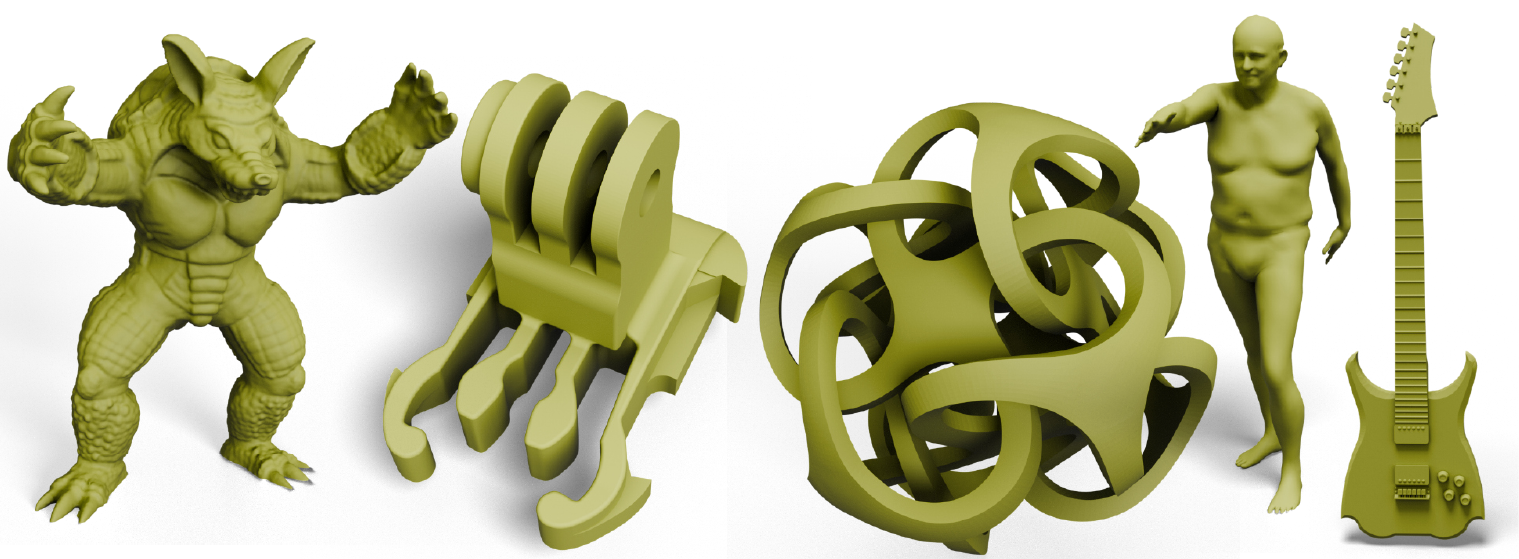}
    \caption{Visualization of samples from our dataset.
    Our dataset covers a large range of mesh complexities, from simple CAD objects composed of a few thousand triangles to highly detailed shapes with millions of triangles.}
    \label{fig:data_sample}
\end{figure}

\begin{figure*}
    \includegraphics[width=\linewidth]{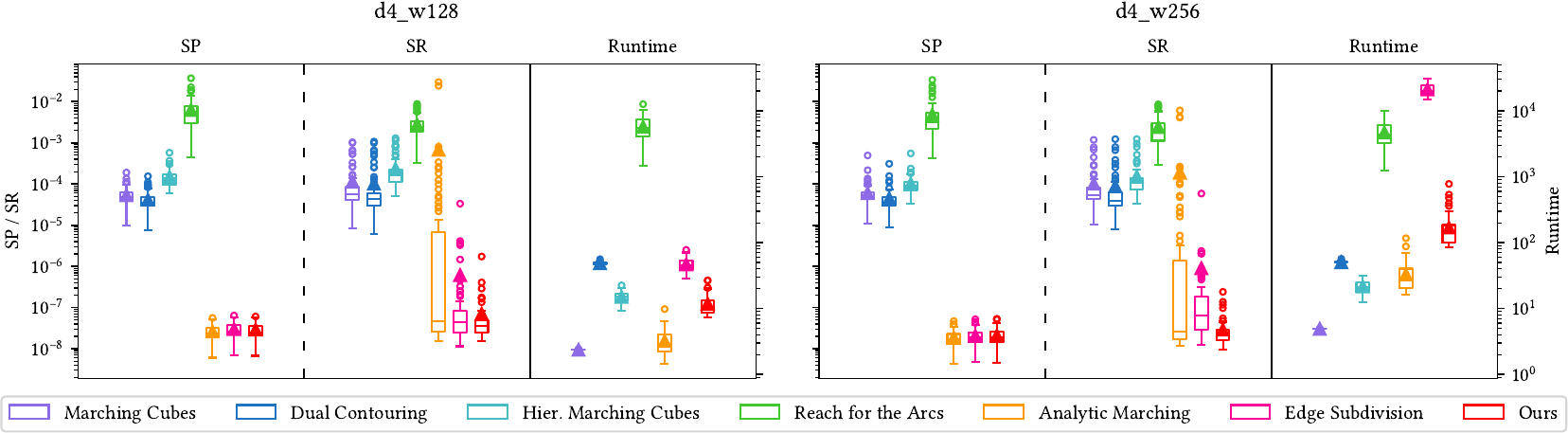}
    \caption{Box plots for soft precision (SP), soft recall (SR), and runtime per architecture and method. Little triangles indicate the means. Runtime is measured in seconds. Most approximation methods use a grid of resolution 512, while \baseline{Reach for the Arcs} uses 128.}
    \label{fig:boxplots}
\end{figure*}

\begin{figure*}
    \includegraphics[width=\linewidth]{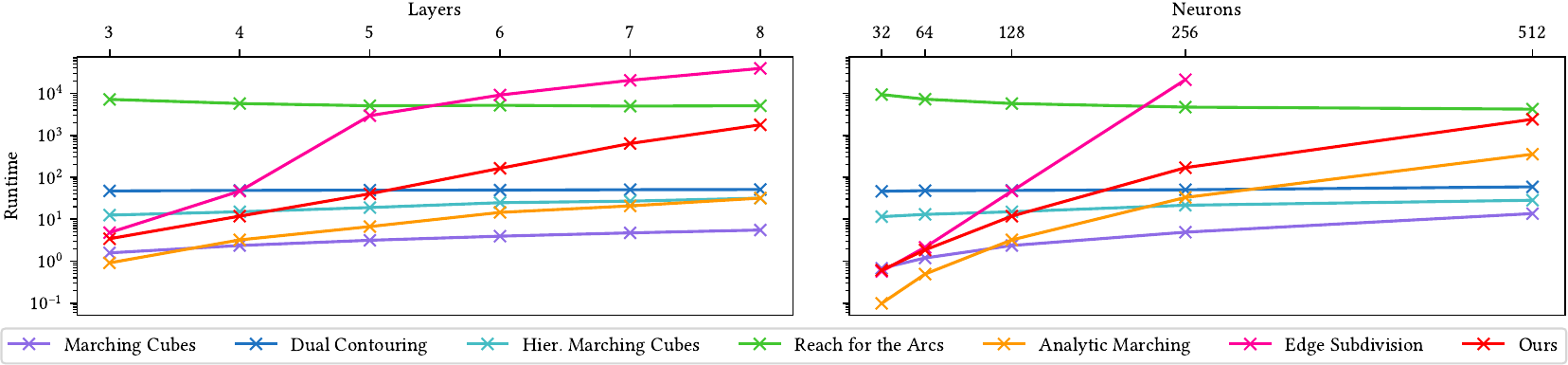}
    \caption{Scalability study with runtime measured in seconds. We vary the depth and width of the \neurarch{d4\_w128} architecture and extract meshes with every method.}
    \label{fig:runtime}
\end{figure*}

We evaluate our method and all other baselines on $84$ watertight shapes collected from five different datasets: $20$ shapes from \dataset{Thingi10K}~\cite{Thingi10K}, $20$ shapes from the \dataset{ABC Dataset}~\cite{Koch_2019_CVPR}, $19$ shapes from \dataset{ShapeNet}~\cite{chang2015shapenet}, $20$ shapes from \dataset{FAUST}~\cite{Bogo_CVPR_FAUST_2014}, and $5$ shapes from the \dataset{Stanford 3D Scanning Repository}~\cite{curless1996volumetric}.
\refFig{data_sample} depicts some samples from our dataset.
Then, for each shape, we fit neural SDFs with all architectures.

\paragraph{SDF Training Protocol.}
Before training, meshes are converted into an SDF.
Each mesh's bounding box is normalized to the range $[-.95,.95]^3$, and then densely sampled to produce $10^7$ training points per shape.
For non-ShapeNet meshes, $2^{20}$ samples ($\approx 6\%$ of the total) are drawn uniformly from the bounding box using a Sobol sequence, while the remaining points ($\approx 94\%$) are concentrated near the surface.
These are split equally into on-surface points and near-surface points (each $\approx 47\%$ of the total).
The near-surface samples are created by adding Gaussian noise with $\sigma=0.05$ to surface points.
Signed distances are computed for the uniform and near-surface samples using Open3D's raycasting~\cite{Zhou2018open3d} with $9$ evaluation rays per query, while the on-surface samples are assigned distance zero.
For ShapeNet meshes, which are often non-watertight, DeepSDF’s dedicated preprocessing tool~\cite{park2019deepsdf} is used to generate the same total number of samples distributed across inside and outside regions.
Networks are trained with the Adam~\cite{kingma2014adam} optimizer at a fixed learning rate of $10^{-4}$ and a batch size of $10^4$.
Batches are sampled with replacement from the full dataset.
Training runs for four hours.

\paragraph{Baselines}

We compare our method to several baselines that approximate the mesh surface by spatial decomposition and sampling of the SDF, and other exact methods like ours.
In particular, we select \baseline{Marching Cubes}~\cite{lorensen1987marching}, \baseline{Dual Contouring}~\cite{ju2002dual}, \baseline{Hierarchical Marching Cubes}~\cite{sharp2022spelunking}, and \baseline{Reach for the Arcs}~\cite{Sellan2024RFTA} as representative baselines for approximation methods.
For such methods, we use a grid resolution of 64, 128, 256 and 512.
Moreover, we compare to two existing exact methods: \baseline{Analytic Marching}~\cite{Lei2021} and \baseline{Edge Subdivision}~\cite{Berzins23}.

\paragraph{Neural Architectures}

We chose a neural architecture commonly used to encode SDFs: a ReLU multi-layer perceptron (MLP).
In particular, we selected two variants: a ReLU MLP with four layers and $128$ neurons each (\neurarch{d4\_w128}), and a more complex architecture also composed of four layers but with $256$ neurons each (\neurarch{d4\_w256}).

\subsection{Main Results}

\refTab{main-results} presents the main result of our comparison.
Naturally, all approximation methods produce meshes with high SP and SR when low-resolution sampling grids are used.
Employing higher-resolution grids decreases these metrics, but leads to an increase in processing time and generated triangles.
Yet, even for the finest grids with resolution $512$, the reconstructed meshes significantly deviate from the SDF.
Moreover, the recent method \baseline{Reach for the Arcs}~\cite{Sellan2024RFTA} is only able to process low-resolution grids due to its long processing times.

Analytic methods, on the other hand, produce meshes with almost perfect SP.
Regarding SR, however, \baseline{Analytic Marching}~\cite{Lei2021} exhibits high metrics as it misses disconnected parts of the mesh.
This is the result of the nature of their algorithm, which relies on a set of seed points from which the mesh is reconstructed iteratively.
In contrast, this is not the case for \baseline{Edge Subdivision}~\cite{Berzins23}, which produces meshes with low SR.
Unfortunately, it takes a long time to reconstruct large networks.
When we look at the distribution of these metrics over the different shapes in our dataset in \refFig{boxplots}, analytic methods present high variability in SR, indicating that they struggle with certain shapes.
Our method, on the other hand, produces more accurate meshes with fewer triangles than other analytic methods, while maintaining a reasonable runtime comparable to approximation methods.

\begin{figure*}
    \includegraphics[width=.95\linewidth]{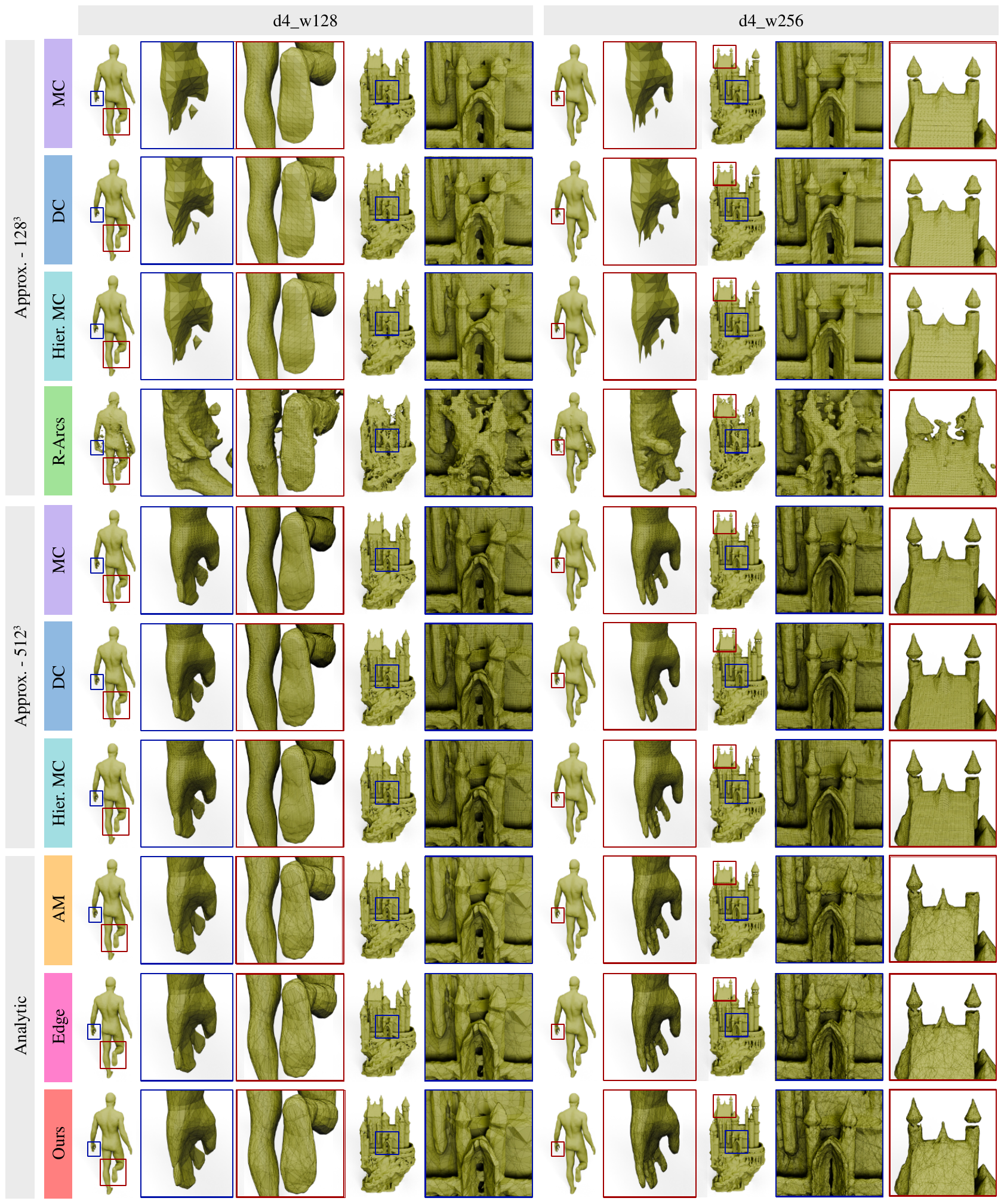}
    \caption{Qualitative results of all the baselines for two network architectures with different numbers of neurons.}
    \label{fig:qualitative}
\end{figure*}

\refFig{qualitative} provides qualitative results of several meshes reconstructed by all baselines.
These show that our method is able to produce more accurate results than approximation methods, which struggle to reconstruct sharp edges, and than \baseline{Analytic Marching}~\cite{Lei2020, Lei2021}, which fails to reconstruct certain disconnected components.
Additionally, \refFig{qualitative} also shows that reconstructions produced by \baseline{Edge Subdivision}~\cite{Berzins23} result in denser meshes with more triangles.

\begin{figure*}
    \includegraphics[width=\linewidth]{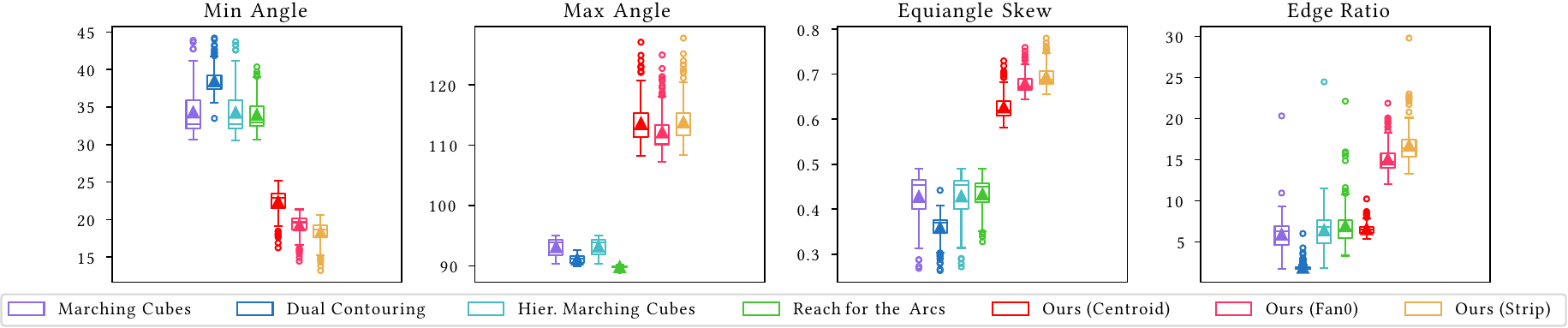}
    \caption{Evaluation of the quality of the triangle meshes generated by different reconstruction methods using the \neurarch{d4\_w256} architecture.}
    \label{fig:triangle_quality}
\end{figure*}

\begin{figure*}
    \includegraphics[width=\linewidth]{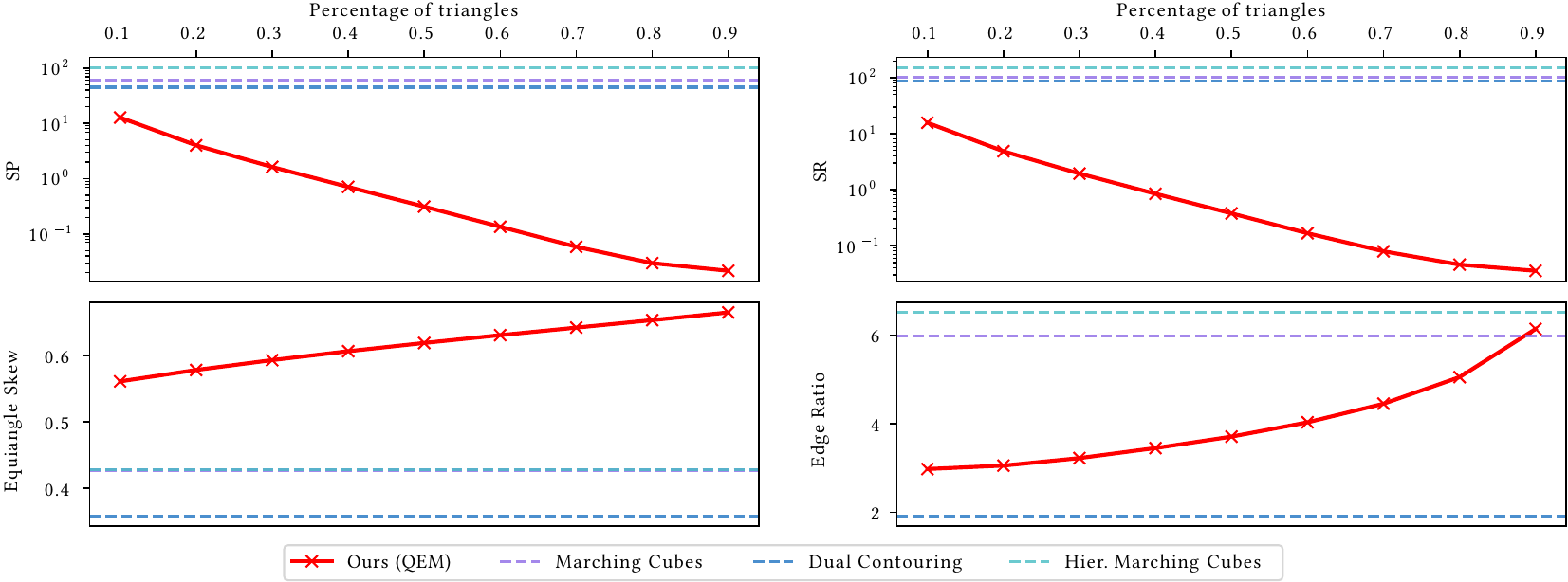}
    \caption{Soft precision (SP), soft recall (SR), and triangle quality metrics of our meshes after reducing the number of triangles using Quadric Error Metrics (QEM). 
    Additionally, for reference, we include results from several approximation methods without any post-processing.
    Note that Marching Cubes and Hier. Marching Cubes overlap on some of the triangle quality metrics.}
    \label{fig:qem_quality}
\end{figure*}

\subsection{Scalability}

We also evaluate how the runtime of each method scales for network architectures with increasing numbers of layers and neurons in \refFig{runtime}.
For most approximation methods, the runtime slightly increases with both the depth and width of the network, since inference becomes more costly. For \baseline{Reach for the Arcs}~\cite{Sellan2024RFTA}, this effect is drowned by the method's generally long runtime.
As expected, analytic methods exhibit a steeper increase in runtime as we increase the network size.
Among them, \baseline{Edge Subdivision}~\cite{Berzins23} is particularly sensitive to the number of neurons, and fails to process a network with width $512$.
On the other hand, both \baseline{Analytic Marching}~\cite{Lei2020, Lei2021} and our method scale more gracefully with the number of neurons, while ours unfortunately scales worse with the number of layers.
Still, these results show that our method is able to process a large range of network architectures in a reasonable runtime similar to \baseline{Analytic Marching}~\cite{Lei2020, Lei2021} while being substantially more accurate.

\subsection{Triangle Mesh Quality}

All analytic extractors output polygonal faces that we tessellate to triangles for downstream use.
In the following experiments, we evaluate the quality of the triangle meshes generated by analytic methods with different tessellation approaches compared to approximation methods.
We report standard triangle-quality metrics.

\paragraph{Tessellation.}
Several strategies were considered for tessellating the $k$-gon faces produced by the analytic extractor.
The \emph{fan$_0$} method connects every triangle to the first vertex $i_0$, yielding the triangle set $\{(i_0,i_j,i_{j+1}) | 1 \leq j \leq k-2\}$. 
However, it often generates long triangles.
\emph{Centroid} introduces the vertex $c=\tfrac{1}{k}\sum_{j=1}^k v_{i_j}$ as an additional vertex. 
Triangles $\{(c,i_j,i_{j+1}) | 0 \leq j \leq k-2\}$ are emitted cyclically, producing a uniform hub structure, though additional triangles are needed.
Lastly, \emph{strip} generates a triangle strip from the polygon, \ie list of triangles created by iteratively generating triangles that share an edge with the previous triangle in the list.

\begin{figure*}
    \includegraphics[width=\linewidth]{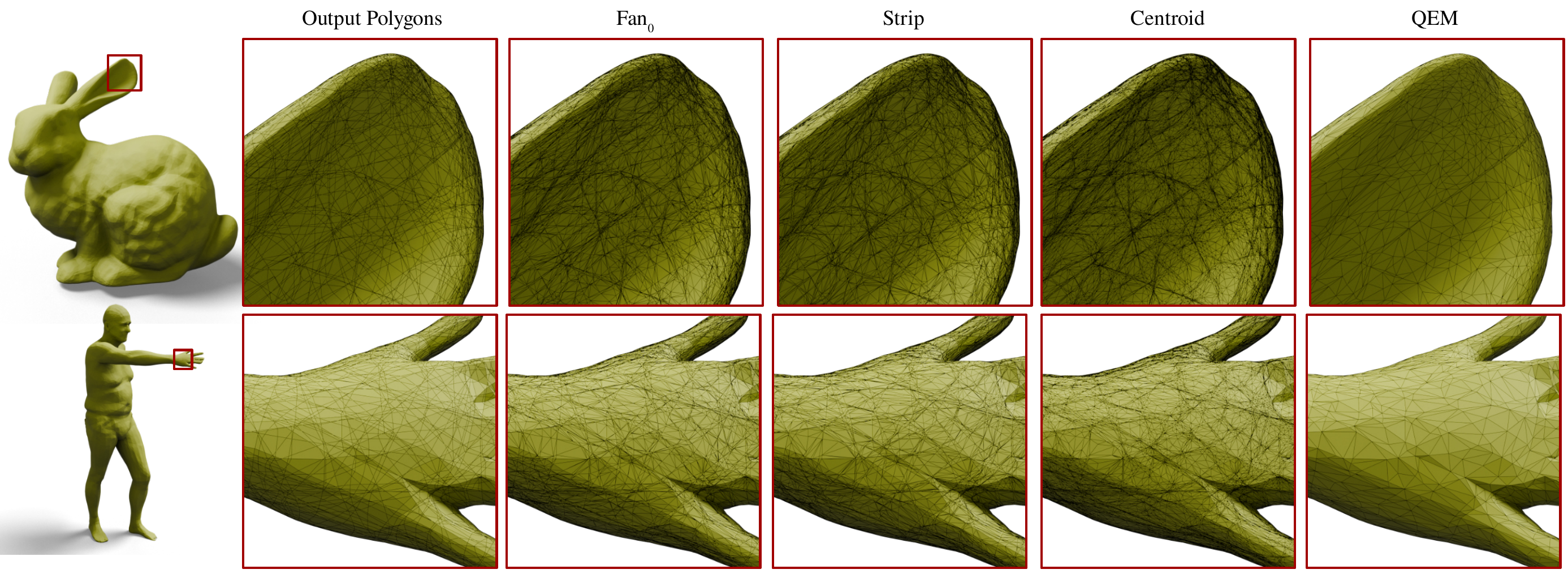}
    \caption{Different approaches of tessellating the reconstructed polygons compared to the simplified mesh containing $10\,\%$ of the original number of triangles.}
    \label{fig:tessellation}
\end{figure*}

\paragraph{Metrics.}
Mesh quality metrics quantify how well a triangulation avoids poorly shaped elements that cause numerical instability.
Good elements have angles close to $60^\circ$, balanced edge lengths, and near-equilateral proportions.
In order to measure such properties, we rely on the following well-established metrics.
We analyze the angles of the generated triangles using the maximum, $\theta_{M}$, and minimum angle, $\theta_{m}$,~\cite{rupert1995delaunay, shewchuk1999lecture} as well as the equiangle skew~\cite{stimpson2007verdict, pebay2007new}, which measures the deviation from an equilateral triangle:
\[
\theta_{s} = \max\!\big((\theta_{M}-60)/120,\,(60-\theta_{m})/60\big)\in[0,1]
\]
Additionally, we use the ratio between the largest and shortest edge, $L_{r}=\ell_M / \ell_m$,~\cite{sorgente2023survey}.

\paragraph{Results.}
\refFig{triangle_quality} presents the results of this experiment.
While all approximation methods provide a similar triangle quality, analytic methods generate triangle meshes with lower quality according to all metrics.
Among the different tessellation approaches, \emph{centroid} provides a slight improvement over \emph{fan$_0$} and \emph{strip} at the cost of additional triangles, with \emph{strip} providing the meshes with the lowest quality.
\refFig{tessellation} presents qualitative results of the different tessellation methods in comparison to the original extracted polygons.

\subsection{Post-processing}
The output meshes from analytic methods are usually composed of many small triangles, which arise from the optimization process.
However, many of these small triangles provide little information about the underlying shape.
Therefore, in this section, we post-process the resulting meshes from our analytic extraction method with a simplification algorithm based on Quadric Error Metrics (QEM)~\cite{garland97QEM}.
\refFig{qem_quality} presents the resulting SP and SR for meshes containing different percentages of the original number of triangles. 
While reducing the number of triangles increases the error of our meshes, those remain smaller than the error produced by all approximation methods, even when we reduce the number of triangles to $10\,\%$ of the original mesh.
Moreover, when we analyze the triangle quality metrics of the simplified meshes, we observe that reducing the number of triangles increases the quality of the triangle mesh, reducing the gap between analytic and approximation methods.
Lastly, \refFig{tessellation} presents some qualitative results of the simplified mesh compared to different tessellation methods, where we observe that the simplified mesh preserves the original shape while drastically reducing the number of small triangles.

\begin{figure}
    \includegraphics[width=\linewidth]{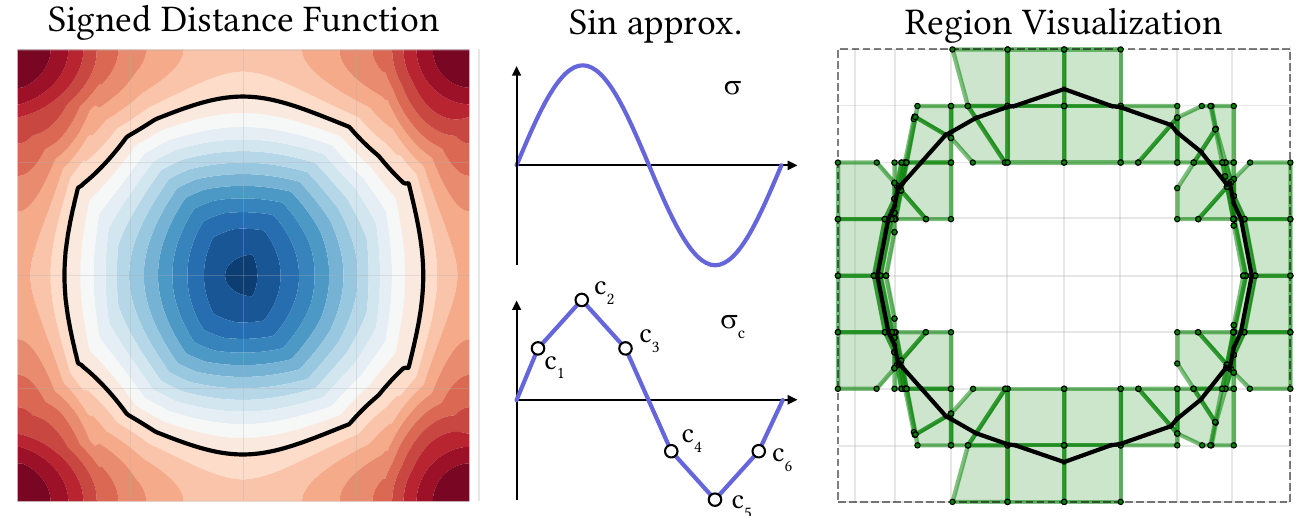}
    \caption{SDF (left) encoded using a neural architecture with positional encoding.
    We use piece-wise linear surrogates for the sinusoids (center) in our reconstruction (right). Green polygons represent the linear regions containing the zero-level set.}
    \label{fig:pe_exp}
\end{figure}

\subsection{Range Analysis Filtering}

Filtering of linear regions using range analysis is a crucial step of our algorithm.
In this section, we evaluate the gains introduced by this step by comparing our marching neurons algorithm to a version of our method where the filtering step is deactivated.
We run both algorithms on SDFs encoded by both of our architectures, \neurarch{d4\_w128}, and \neurarch{d4\_w256}.
Unfortunately, for the \neurarch{d4\_w256} architecture, the version without filtering was not able to finish in a window time of two hours for a single mesh due to the exponential grow of the number of linear regions.
For the \neurarch{d4\_w128} architecture, the filtering step lead to an average speed up of $\times 12.2$, highlighting the importance of this step.

\subsection{Extension to Other Activations}

Our method was designed for neural network architectures with ReLU activation functions, but it can easily generalize to piecewise-linear activations such as leaky ReLU by storing slope and intercept instead of a binary mask.
Recent implicit neural representations rely on continuous activation functions such as sine and cosine \cite{tancik2020fourier,sitzmann2019siren} to better capture high-frequency details of the surface.
In order to use our method with these architectures, such activation functions should be approximated with piecewise-linear surrogates.

To evaluate the viability of such an approach, we train a small ReLU network with two layers and four neurons each to approximate the SDF of a circle in 2D.
In this architecture, we process the input coordinates with a positional encoding layer with two frequencies~\cite{mildenhall2021nerf}
During reconstruction, we approximate all periodic functions of the positional encoding layer using a piece-wise linear function composed using 6 knots per period, see \refFig{pe_exp} right.
\refFig{pe_exp} also presents the result of such experiment, where our method is able to reconstruct the underlying SDF via the piece-wise linear approximations.
Unfortunately, our method introduces certain errors in the reconstruction.
The magnitude of this error is defined by the number of linear segments used in the approximation, visible as the square pattern of the linear regions in \refFig{pe_exp}.
However, this error can be reduced by increasing the number of linear segments of the piece-wise linear surrogate.


\section{Limitations}

Our method is not exempt from limitations.
Notably, we rely on range analysis~\cite{duff1992interval, rump2015implementation} to reduce the number of linear cells processed.
If the bounds yielded by this analysis are too conservative, or the SDF encodes a complex shape with little empty space, our method needs to retain a lot of cells, leading to increased memory usage and long reconstruction time.


\section{Conclusion}

We have introduced a novel method for analytic surface extraction from neural implicit shapes that achieves unprecedented accuracy.
This was achieved through a natively parallel algorithm design that combines recursive polygon splitting with range analysis to filter empty regions.
Our meshes accurately capture the full detail encoded in the neural implicit function by departing from the common practice of treating neural networks as black boxes.

These properties open up promising avenues for future work. 
Our recursive subdivision scheme is likely to be well-suited for level-of-detail generation, offering greater flexibility for subsequent processing steps that require lower-resolution meshes.
Finally, embedding our mesh extractor into an end-to-end differentiable pipeline would enable optimizing a neural SDF with mesh-based supervision, thus unlocking the potential of our approach on a wide range of tasks.

\begin{acks}
This work was partially funded by the Austrian Research Promotion Agency (FFG) under the project ``AUTARK – Energy-Aware AMR Safety'' (No. 999922723), within the call ``Key Technologies for the Future – National 2024''. We gratefully acknowledge Wolfgang Koch for the insightful and valuable research discussions.
\end{acks}


\newpage
\bibliographystyle{ACM-Reference-Format}
\bibliography{bib}


\begin{thebibliography}{77}


\ifx \showCODEN    \undefined \def \showCODEN     #1{\unskip}     \fi
\ifx \showISBNx    \undefined \def \showISBNx     #1{\unskip}     \fi
\ifx \showISBNxiii \undefined \def \showISBNxiii  #1{\unskip}     \fi
\ifx \showISSN     \undefined \def \showISSN      #1{\unskip}     \fi
\ifx \showLCCN     \undefined \def \showLCCN      #1{\unskip}     \fi
\ifx \shownote     \undefined \def \shownote      #1{#1}          \fi
\ifx \showarticletitle \undefined \def \showarticletitle #1{#1}   \fi
\ifx \showURL      \undefined \def \showURL       {\relax}        \fi
\providecommand\bibfield[2]{#2}
\providecommand\bibinfo[2]{#2}
\providecommand\natexlab[1]{#1}
\providecommand\showeprint[2][]{arXiv:#2}

\bibitem[Akenine-Moller et~al\mbox{.}(2019)]%
        {akenine2019real}
\bibfield{author}{\bibinfo{person}{Tomas Akenine-Moller}, \bibinfo{person}{Eric
  Haines}, {and} \bibinfo{person}{Naty Hoffman}.}
  \bibinfo{year}{2019}\natexlab{}.
\newblock \bibinfo{booktitle}{\emph{Real-time rendering}}.
\newblock \bibinfo{publisher}{AK Peters/crc Press}.
\newblock


\bibitem[Azernikov and Fischer(2005)]%
        {azernikov2005anisotropic}
\bibfield{author}{\bibinfo{person}{Sergei Azernikov} {and}
  \bibinfo{person}{Anath Fischer}.} \bibinfo{year}{2005}\natexlab{}.
\newblock \showarticletitle{Anisotropic meshing of implicit surfaces}. In
  \bibinfo{booktitle}{\emph{International Conference on Shape Modeling and
  Applications}}. \bibinfo{pages}{94--103}.
\newblock


\bibitem[Berzins(2023)]%
        {Berzins23}
\bibfield{author}{\bibinfo{person}{Arturs Berzins}.}
  \bibinfo{year}{2023}\natexlab{}.
\newblock \showarticletitle{Polyhedral Complex Extraction from ReLU Networks
  using Edge Subdivision}. In \bibinfo{booktitle}{\emph{International
  Conference on Machine Learning (ICML)}}.
\newblock


\bibitem[Bloomenthal(1988)]%
        {bloomenthal1988polygonization}
\bibfield{author}{\bibinfo{person}{Jules Bloomenthal}.}
  \bibinfo{year}{1988}\natexlab{}.
\newblock \showarticletitle{Polygonization of implicit surfaces}.
\newblock \bibinfo{journal}{\emph{Computer Aided Geometric Design}}
  \bibinfo{volume}{5}, \bibinfo{number}{4} (\bibinfo{year}{1988}),
  \bibinfo{pages}{341--355}.
\newblock


\bibitem[Bloomenthal et~al\mbox{.}(1997)]%
        {bloomenthal1997introduction}
\bibfield{author}{\bibinfo{person}{Jules Bloomenthal}, \bibinfo{person}{Brian
  Wyvill}, \bibinfo{person}{Geoff Wyvill}, \bibinfo{person}{Alan~H. Barr},
  {and} \bibinfo{person}{Alyn~P. Rockwood}.} \bibinfo{year}{1997}\natexlab{}.
\newblock \bibinfo{booktitle}{\emph{Introduction to Implicit Surfaces}}.
\newblock


\bibitem[Bogo et~al\mbox{.}(2014)]%
        {Bogo_CVPR_FAUST_2014}
\bibfield{author}{\bibinfo{person}{Federica Bogo}, \bibinfo{person}{Javier
  Romero}, \bibinfo{person}{Matthew Loper}, {and} \bibinfo{person}{Michael~J.
  Black}.} \bibinfo{year}{2014}\natexlab{}.
\newblock \showarticletitle{{FAUST}: Dataset and evaluation for {3D} mesh
  registration}. In \bibinfo{booktitle}{\emph{IEEE/CVF Conference on Computer
  Vision and Pattern Recognition (CVPR)}}.
\newblock


\bibitem[Botsch et~al\mbox{.}(2010)]%
        {botsch2010polygon}
\bibfield{author}{\bibinfo{person}{Mario Botsch}, \bibinfo{person}{Leif
  Kobbelt}, \bibinfo{person}{Mark Pauly}, \bibinfo{person}{Pierre Alliez},
  {and} \bibinfo{person}{Bruno L{\'e}vy}.} \bibinfo{year}{2010}\natexlab{}.
\newblock \bibinfo{booktitle}{\emph{Polygon mesh processing}}.
\newblock \bibinfo{publisher}{CRC press}.
\newblock


\bibitem[Bradbury et~al\mbox{.}(2018)]%
        {bradbury2018jax}
\bibfield{author}{\bibinfo{person}{James Bradbury}, \bibinfo{person}{Roy
  Frostig}, \bibinfo{person}{Peter Hawkins}, \bibinfo{person}{Matthew~James
  Johnson}, \bibinfo{person}{Chris Leary}, \bibinfo{person}{Dougal Maclaurin},
  \bibinfo{person}{George Necula}, \bibinfo{person}{Adam Paszke},
  \bibinfo{person}{Jake VanderPlas}, \bibinfo{person}{Skye Wanderman-Milne},
  {and} \bibinfo{person}{Qiao Zhang}.} \bibinfo{year}{2018}\natexlab{}.
\newblock \showarticletitle{JAX: {C}omposable transformations of Python+ NumPy
  programs}.
\newblock  (\bibinfo{year}{2018}).
\newblock


\bibitem[Chang et~al\mbox{.}(2015)]%
        {chang2015shapenet}
\bibfield{author}{\bibinfo{person}{Angel~X Chang}, \bibinfo{person}{Thomas
  Funkhouser}, \bibinfo{person}{Leonidas Guibas}, \bibinfo{person}{Pat
  Hanrahan}, \bibinfo{person}{Qixing Huang}, \bibinfo{person}{Zimo Li},
  \bibinfo{person}{Silvio Savarese}, \bibinfo{person}{Manolis Savva},
  \bibinfo{person}{Shuran Song}, \bibinfo{person}{Hao Su}, {et~al\mbox{.}}}
  \bibinfo{year}{2015}\natexlab{}.
\newblock \showarticletitle{ShapeNet: An Information-Rich 3D Model Repository}.
\newblock \bibinfo{journal}{\emph{arXiv preprint arXiv:1512.03012}}
  (\bibinfo{year}{2015}).
\newblock


\bibitem[Chen et~al\mbox{.}(2022)]%
        {chen2022neural}
\bibfield{author}{\bibinfo{person}{Zhiqin Chen}, \bibinfo{person}{Andrea
  Tagliasacchi}, \bibinfo{person}{Thomas Funkhouser}, {and}
  \bibinfo{person}{Hao Zhang}.} \bibinfo{year}{2022}\natexlab{}.
\newblock \showarticletitle{Neural dual contouring}.
\newblock \bibinfo{journal}{\emph{ACM Transactions on Graphics (TOG)}}
  \bibinfo{volume}{41}, \bibinfo{number}{4} (\bibinfo{year}{2022}),
  \bibinfo{pages}{1--13}.
\newblock


\bibitem[Chen and Zhang(2021)]%
        {chen2021neural}
\bibfield{author}{\bibinfo{person}{Zhiqin Chen} {and} \bibinfo{person}{Hao
  Zhang}.} \bibinfo{year}{2021}\natexlab{}.
\newblock \showarticletitle{Neural marching cubes}.
\newblock \bibinfo{journal}{\emph{ACM Transactions on Graphics (TOG)}}
  \bibinfo{volume}{40}, \bibinfo{number}{6} (\bibinfo{year}{2021}),
  \bibinfo{pages}{1--15}.
\newblock


\bibitem[Chernyaev(1995)]%
        {chernyaev1995marching}
\bibfield{author}{\bibinfo{person}{Evgeni Chernyaev}.}
  \bibinfo{year}{1995}\natexlab{}.
\newblock \bibinfo{booktitle}{\emph{Marching cubes 33: Construction of
  topologically correct isosurfaces}}.
\newblock \bibinfo{type}{{T}echnical {R}eport}.
\newblock


\bibitem[Chibane et~al\mbox{.}(2020)]%
        {chibane2020neural}
\bibfield{author}{\bibinfo{person}{Julian Chibane}, \bibinfo{person}{Aymen
  Mir}, {and} \bibinfo{person}{Gerard Pons-Moll}.}
  \bibinfo{year}{2020}\natexlab{}.
\newblock \showarticletitle{Neural Unsigned Distance Fields for Implicit
  Function Learning}. In \bibinfo{booktitle}{\emph{Advances in Neural
  Information Processing Systems (NeurIPS)}}.
\newblock


\bibitem[Comba and Stolfi(1993)]%
        {ld1993affine}
\bibfield{author}{\bibinfo{person}{Joao L.~D. Comba} {and} \bibinfo{person}{J.
  Stolfi}.} \bibinfo{year}{1993}\natexlab{}.
\newblock \showarticletitle{Affine arithmetic and its applications to computer
  graphics}.
\newblock \bibinfo{journal}{\emph{SIBGRAPI'93}} (\bibinfo{year}{1993}).
\newblock


\bibitem[Curless and Levoy(1996)]%
        {curless1996volumetric}
\bibfield{author}{\bibinfo{person}{Brian Curless} {and} \bibinfo{person}{Marc
  Levoy}.} \bibinfo{year}{1996}\natexlab{}.
\newblock \showarticletitle{A volumetric method for building complex models
  from range images}. In \bibinfo{booktitle}{\emph{Conference on Computer
  Graphics and Interactive Techniques}}. \bibinfo{pages}{303--312}.
\newblock


\bibitem[De~Ara{\'u}jo et~al\mbox{.}(2015)]%
        {de2015survey}
\bibfield{author}{\bibinfo{person}{Bruno~Rodrigues De~Ara{\'u}jo},
  \bibinfo{person}{Daniel~S Lopes}, \bibinfo{person}{Pauline Jepp},
  \bibinfo{person}{Joaquim~A Jorge}, {and} \bibinfo{person}{Brian Wyvill}.}
  \bibinfo{year}{2015}\natexlab{}.
\newblock \showarticletitle{A survey on implicit surface polygonization}.
\newblock \bibinfo{journal}{\emph{ACM Computing Surveys (CSUR)}}
  \bibinfo{volume}{47}, \bibinfo{number}{4} (\bibinfo{year}{2015}),
  \bibinfo{pages}{1--39}.
\newblock


\bibitem[De~Berg(2000)]%
        {de2000computational}
\bibfield{author}{\bibinfo{person}{Mark De~Berg}.}
  \bibinfo{year}{2000}\natexlab{}.
\newblock \bibinfo{booktitle}{\emph{Computational geometry: algorithms and
  applications}}.
\newblock \bibinfo{publisher}{Springer Science \& Business Media}.
\newblock


\bibitem[Doi and Koide(1991)]%
        {doi1991efficient}
\bibfield{author}{\bibinfo{person}{Akio Doi} {and} \bibinfo{person}{Akio
  Koide}.} \bibinfo{year}{1991}\natexlab{}.
\newblock \showarticletitle{An efficient method of triangulating equi-valued
  surfaces by using tetrahedral cells}.
\newblock \bibinfo{journal}{\emph{IEICE Transactions on Information and
  Systems}} \bibinfo{volume}{74}, \bibinfo{number}{1} (\bibinfo{year}{1991}),
  \bibinfo{pages}{214--224}.
\newblock


\bibitem[Duff(1992)]%
        {duff1992interval}
\bibfield{author}{\bibinfo{person}{Tom Duff}.} \bibinfo{year}{1992}\natexlab{}.
\newblock \showarticletitle{Interval arithmetic recursive subdivision for
  implicit functions and constructive solid geometry}.
\newblock \bibinfo{journal}{\emph{ACM SIGGRAPH Computer Graphics}}
  \bibinfo{volume}{26}, \bibinfo{number}{2} (\bibinfo{year}{1992}),
  \bibinfo{pages}{131--138}.
\newblock


\bibitem[Garland and Heckbert(1997)]%
        {garland97QEM}
\bibfield{author}{\bibinfo{person}{Michael Garland} {and}
  \bibinfo{person}{Paul~S. Heckbert}.} \bibinfo{year}{1997}\natexlab{}.
\newblock \bibinfo{booktitle}{\emph{Surface Simplification Using Quadric Error
  Metrics}}.
\newblock


\bibitem[Goodfellow et~al\mbox{.}(2016)]%
        {Goodfellow-et-al-2016}
\bibfield{author}{\bibinfo{person}{Ian Goodfellow}, \bibinfo{person}{Yoshua
  Bengio}, {and} \bibinfo{person}{Aaron Courville}.}
  \bibinfo{year}{2016}\natexlab{}.
\newblock \bibinfo{booktitle}{\emph{Deep Learning}}.
\newblock \bibinfo{publisher}{MIT Press}.
\newblock
\newblock
\shownote{\url{http://www.deeplearningbook.org}}.


\bibitem[Grigsby and Lindsey(2022)]%
        {grigsby2022transversality}
\bibfield{author}{\bibinfo{person}{J~Elisenda Grigsby} {and}
  \bibinfo{person}{Kathryn Lindsey}.} \bibinfo{year}{2022}\natexlab{}.
\newblock \showarticletitle{On transversality of bent hyperplane arrangements
  and the topological expressiveness of ReLU neural networks}.
\newblock \bibinfo{journal}{\emph{SIAM Journal on Applied Algebra and
  Geometry}} \bibinfo{volume}{6}, \bibinfo{number}{2} (\bibinfo{year}{2022}),
  \bibinfo{pages}{216--242}.
\newblock


\bibitem[Hanin and Rolnick(2019)]%
        {hanin2019deep}
\bibfield{author}{\bibinfo{person}{Boris Hanin} {and} \bibinfo{person}{David
  Rolnick}.} \bibinfo{year}{2019}\natexlab{}.
\newblock \showarticletitle{Deep relu networks have surprisingly few activation
  patterns}.
\newblock \bibinfo{journal}{\emph{Advances in Neural Information Processing
  Systems (NeurIPS)}}  \bibinfo{volume}{32} (\bibinfo{year}{2019}).
\newblock


\bibitem[Hanocka et~al\mbox{.}(2020)]%
        {hanocka2020point2mesh}
\bibfield{author}{\bibinfo{person}{Rana Hanocka}, \bibinfo{person}{Gal Metzer},
  \bibinfo{person}{Raja Giryes}, {and} \bibinfo{person}{Daniel Cohen-Or}.}
  \bibinfo{year}{2020}\natexlab{}.
\newblock \showarticletitle{Point2Mesh: a self-prior for deformable meshes}.
\newblock \bibinfo{journal}{\emph{ACM Transactions on Graphics (TOG)}}
  \bibinfo{volume}{39}, \bibinfo{number}{4} (\bibinfo{year}{2020}),
  \bibinfo{pages}{126--1}.
\newblock


\bibitem[Hege et~al\mbox{.}(1997)]%
        {hege1997generalized}
\bibfield{author}{\bibinfo{person}{Hans-Christian Hege},
  \bibinfo{person}{Martin Seebass}, \bibinfo{person}{Detlev Stalling}, {and}
  \bibinfo{person}{Malte Z{\"o}ckler}.} \bibinfo{year}{1997}\natexlab{}.
\newblock \showarticletitle{A generalized marching cubes algorithm based on
  non-binary classifications}.
\newblock  (\bibinfo{year}{1997}).
\newblock


\bibitem[Hein et~al\mbox{.}(2019)]%
        {hein2019relu}
\bibfield{author}{\bibinfo{person}{Matthias Hein}, \bibinfo{person}{Maksym
  Andriushchenko}, {and} \bibinfo{person}{Julian Bitterwolf}.}
  \bibinfo{year}{2019}\natexlab{}.
\newblock \showarticletitle{Why relu networks yield high-confidence predictions
  far away from the training data and how to mitigate the problem}. In
  \bibinfo{booktitle}{\emph{IEEE/CVF Conference on Computer Vision and Pattern
  Recognition (CVPR)}}. \bibinfo{pages}{41--50}.
\newblock


\bibitem[Hilton et~al\mbox{.}(1996)]%
        {hilton1996marching}
\bibfield{author}{\bibinfo{person}{Adrian Hilton}, \bibinfo{person}{Andrew~J
  Stoddart}, \bibinfo{person}{John Illingworth}, {and} \bibinfo{person}{Terry
  Windeatt}.} \bibinfo{year}{1996}\natexlab{}.
\newblock \showarticletitle{Marching triangles: range image fusion for complex
  object modelling}. In \bibinfo{booktitle}{\emph{IEEE International Conference
  on Image Processing (ICIP)}}, Vol.~\bibinfo{volume}{2}.
  \bibinfo{pages}{381--384}.
\newblock


\bibitem[Hornik et~al\mbox{.}(1989)]%
        {hornik1989multilayer}
\bibfield{author}{\bibinfo{person}{Kurt Hornik}, \bibinfo{person}{Maxwell
  Stinchcombe}, {and} \bibinfo{person}{Halbert White}.}
  \bibinfo{year}{1989}\natexlab{}.
\newblock \showarticletitle{Multilayer feedforward networks are universal
  approximators}.
\newblock \bibinfo{journal}{\emph{Neural networks}} \bibinfo{volume}{2},
  \bibinfo{number}{5} (\bibinfo{year}{1989}), \bibinfo{pages}{359--366}.
\newblock


\bibitem[Humayun et~al\mbox{.}(2023)]%
        {Humayun_2023_CVPR}
\bibfield{author}{\bibinfo{person}{Ahmed~Imtiaz Humayun},
  \bibinfo{person}{Randall Balestriero}, \bibinfo{person}{Guha Balakrishnan},
  {and} \bibinfo{person}{Richard~G. Baraniuk}.}
  \bibinfo{year}{2023}\natexlab{}.
\newblock \showarticletitle{SplineCam: Exact Visualization and Characterization
  of Deep Network Geometry and Decision Boundaries}. In
  \bibinfo{booktitle}{\emph{IEEE/CVF Conference on Computer Vision and Pattern
  Recognition (CVPR)}}. \bibinfo{pages}{3789--3798}.
\newblock


\bibitem[Ju et~al\mbox{.}(2002)]%
        {ju2002dual}
\bibfield{author}{\bibinfo{person}{Tao Ju}, \bibinfo{person}{Frank Losasso},
  \bibinfo{person}{Scott Schaefer}, {and} \bibinfo{person}{Joe Warren}.}
  \bibinfo{year}{2002}\natexlab{}.
\newblock \showarticletitle{Dual contouring of hermite data}. In
  \bibinfo{booktitle}{\emph{Conference on Computer Graphics and Interactive
  Techniques}}. \bibinfo{pages}{339--346}.
\newblock


\bibitem[Kingma and Ba(2015)]%
        {kingma2014adam}
\bibfield{author}{\bibinfo{person}{Diederik~P. Kingma} {and}
  \bibinfo{person}{Jimmy Ba}.} \bibinfo{year}{2015}\natexlab{}.
\newblock \showarticletitle{Adam: {A} Method for Stochastic Optimization}. In
  \bibinfo{booktitle}{\emph{International Conference on Learning
  Representations (ICLR)}}.
\newblock


\bibitem[Koch et~al\mbox{.}(2019)]%
        {Koch_2019_CVPR}
\bibfield{author}{\bibinfo{person}{Sebastian Koch}, \bibinfo{person}{Albert
  Matveev}, \bibinfo{person}{Zhongshi Jiang}, \bibinfo{person}{Francis
  Williams}, \bibinfo{person}{Alexey Artemov}, \bibinfo{person}{Evgeny
  Burnaev}, \bibinfo{person}{Marc Alexa}, \bibinfo{person}{Denis Zorin}, {and}
  \bibinfo{person}{Daniele Panozzo}.} \bibinfo{year}{2019}\natexlab{}.
\newblock \showarticletitle{ABC: A Big CAD Model Dataset For Geometric Deep
  Learning}. In \bibinfo{booktitle}{\emph{IEEE/CVF Conference on Computer
  Vision and Pattern Recognition (CVPR)}}.
\newblock


\bibitem[Kohlbrenner and Alexa(2025)]%
        {kohlbrenner25}
\bibfield{author}{\bibinfo{person}{M. Kohlbrenner} {and} \bibinfo{person}{M.
  Alexa}.} \bibinfo{year}{2025}\natexlab{}.
\newblock \showarticletitle{Isosurface Extraction for Signed Distance Functions
  using Power Diagrams}.
\newblock \bibinfo{journal}{\emph{Computer Graphics Forum}}
  (\bibinfo{year}{2025}).
\newblock


\bibitem[Kunc and Kl{\'e}ma(2024)]%
        {kunc2024three}
\bibfield{author}{\bibinfo{person}{Vladim{\'\i}r Kunc} {and}
  \bibinfo{person}{Ji{\v{r}}{\'\i} Kl{\'e}ma}.}
  \bibinfo{year}{2024}\natexlab{}.
\newblock \showarticletitle{Three Decades of Activations: A Comprehensive
  Survey of 400 Activation Functions for Neural Networks}.
\newblock \bibinfo{journal}{\emph{arXiv preprint arXiv:2402.09092}}
  (\bibinfo{year}{2024}).
\newblock


\bibitem[Lei and Jia(2020)]%
        {Lei2020}
\bibfield{author}{\bibinfo{person}{Jiabao Lei} {and} \bibinfo{person}{Kui
  Jia}.} \bibinfo{year}{2020}\natexlab{}.
\newblock \showarticletitle{Analytic Marching: An Analytic Meshing Solution
  from Deep Implicit Surface Networks}. In
  \bibinfo{booktitle}{\emph{International Conference on Machine Learning
  (ICML)}}.
\newblock


\bibitem[Lei et~al\mbox{.}(2021)]%
        {Lei2021}
\bibfield{author}{\bibinfo{person}{Jiabao Lei}, \bibinfo{person}{Kui Jia},
  {and} \bibinfo{person}{Yi Ma}.} \bibinfo{year}{2021}\natexlab{}.
\newblock \showarticletitle{Learning and Meshing from Deep Implicit Surface
  Networks Using an Efficient Implementation of Analytic Marching}.
\newblock \bibinfo{journal}{\emph{IEEE Transactions on Pattern Analysis and
  Machine Intelligence (TPAMI)}} (\bibinfo{year}{2021}), \bibinfo{pages}{1--1}.
\newblock


\bibitem[Lorensen and Cline(1987)]%
        {lorensen1987marching}
\bibfield{author}{\bibinfo{person}{William~E Lorensen} {and}
  \bibinfo{person}{Harvey~E Cline}.} \bibinfo{year}{1987}\natexlab{}.
\newblock \showarticletitle{Marching cubes: A high resolution 3D surface
  construction algorithm}.
\newblock \bibinfo{journal}{\emph{ACM SIGGRAPH Computer Graphics}}
  \bibinfo{volume}{21}, \bibinfo{number}{4} (\bibinfo{year}{1987}),
  \bibinfo{pages}{163--169}.
\newblock


\bibitem[Marschner et~al\mbox{.}(2023)]%
        {marschner2023constructive}
\bibfield{author}{\bibinfo{person}{Zo{\"e} Marschner}, \bibinfo{person}{Silvia
  Sell{\'a}n}, \bibinfo{person}{Hsueh-Ti~Derek Liu}, {and}
  \bibinfo{person}{Alec Jacobson}.} \bibinfo{year}{2023}\natexlab{}.
\newblock \showarticletitle{Constructive solid geometry on neural signed
  distance fields}. In \bibinfo{booktitle}{\emph{SIGGRAPH Asia}}.
  \bibinfo{pages}{1--12}.
\newblock


\bibitem[Mildenhall et~al\mbox{.}(2020)]%
        {mildenhall2021nerf}
\bibfield{author}{\bibinfo{person}{Ben Mildenhall}, \bibinfo{person}{Pratul~P.
  Srinivasan}, \bibinfo{person}{Matthew Tancik}, \bibinfo{person}{Jonathan~T.
  Barron}, \bibinfo{person}{Ravi Ramamoorthi}, {and} \bibinfo{person}{Ren Ng}.}
  \bibinfo{year}{2020}\natexlab{}.
\newblock \showarticletitle{NeRF: Representing Scenes as Neural Radiance Fields
  for View Synthesis}. In \bibinfo{booktitle}{\emph{European Conference on
  Computer Vision (ECCV)}}.
\newblock


\bibitem[Montani et~al\mbox{.}(1994)]%
        {montani1994discretized}
\bibfield{author}{\bibinfo{person}{Claudio Montani}, \bibinfo{person}{Riccardo
  Scateni}, {and} \bibinfo{person}{Roberto Scopigno}.}
  \bibinfo{year}{1994}\natexlab{}.
\newblock \showarticletitle{Discretized marching cubes}. In
  \bibinfo{booktitle}{\emph{Proceedings Visualization}}.
  \bibinfo{pages}{281--287}.
\newblock


\bibitem[Montufar et~al\mbox{.}(2014)]%
        {montufar2014number}
\bibfield{author}{\bibinfo{person}{Guido~F Montufar}, \bibinfo{person}{Razvan
  Pascanu}, \bibinfo{person}{Kyunghyun Cho}, {and} \bibinfo{person}{Yoshua
  Bengio}.} \bibinfo{year}{2014}\natexlab{}.
\newblock \showarticletitle{On the number of linear regions of deep neural
  networks}.
\newblock \bibinfo{journal}{\emph{Advances in Neural Information Processing
  Systems (NeurIPS)}}  \bibinfo{volume}{27} (\bibinfo{year}{2014}).
\newblock


\bibitem[Naitzat et~al\mbox{.}(2020)]%
        {naitzat2020topology}
\bibfield{author}{\bibinfo{person}{Gregory Naitzat}, \bibinfo{person}{Andrey
  Zhitnikov}, {and} \bibinfo{person}{Lek-Heng Lim}.}
  \bibinfo{year}{2020}\natexlab{}.
\newblock \showarticletitle{Topology of deep neural networks}.
\newblock \bibinfo{journal}{\emph{Journal of Machine Learning Research}}
  \bibinfo{volume}{21}, \bibinfo{number}{184} (\bibinfo{year}{2020}),
  \bibinfo{pages}{1--40}.
\newblock


\bibitem[Newman and Yi(2006)]%
        {NEWMAN2006854}
\bibfield{author}{\bibinfo{person}{Timothy~S. Newman} {and}
  \bibinfo{person}{Hong Yi}.} \bibinfo{year}{2006}\natexlab{}.
\newblock \showarticletitle{A survey of the marching cubes algorithm}.
\newblock \bibinfo{journal}{\emph{Computers \& Graphics}} \bibinfo{volume}{30},
  \bibinfo{number}{5} (\bibinfo{year}{2006}), \bibinfo{pages}{854--879}.
\newblock
\showISSN{0097-8493}


\bibitem[Nielson(2004)]%
        {nielson2004dual}
\bibfield{author}{\bibinfo{person}{Gregory~M Nielson}.}
  \bibinfo{year}{2004}\natexlab{}.
\newblock \showarticletitle{Dual marching cubes}. In
  \bibinfo{booktitle}{\emph{Visualization}}. \bibinfo{pages}{489--496}.
\newblock


\bibitem[Osher et~al\mbox{.}(2004)]%
        {osher2004level}
\bibfield{author}{\bibinfo{person}{Stanley Osher}, \bibinfo{person}{Ronald
  Fedkiw}, {and} \bibinfo{person}{K Piechor}.} \bibinfo{year}{2004}\natexlab{}.
\newblock \showarticletitle{Level set methods and dynamic implicit surfaces}.
\newblock \bibinfo{journal}{\emph{Appl. Mech. Rev.}} \bibinfo{volume}{57},
  \bibinfo{number}{3} (\bibinfo{year}{2004}), \bibinfo{pages}{B15--B15}.
\newblock


\bibitem[Park et~al\mbox{.}(2019)]%
        {park2019deepsdf}
\bibfield{author}{\bibinfo{person}{Jeong~Joon Park}, \bibinfo{person}{Peter
  Florence}, \bibinfo{person}{Julian Straub}, \bibinfo{person}{Richard
  Newcombe}, {and} \bibinfo{person}{Steven Lovegrove}.}
  \bibinfo{year}{2019}\natexlab{}.
\newblock \showarticletitle{Deep{SDF}: Learning continuous signed distance
  functions for shape representation}. In \bibinfo{booktitle}{\emph{IEEE/CVF
  Conference on Computer Vision and Pattern Recognition (CVPR)}}.
  \bibinfo{pages}{165--174}.
\newblock


\bibitem[Pascanu et~al\mbox{.}(2014)]%
        {pascanu2013number}
\bibfield{author}{\bibinfo{person}{Razvan Pascanu}, \bibinfo{person}{Guido
  Montufar}, {and} \bibinfo{person}{Yoshua Bengio}.}
  \bibinfo{year}{2014}\natexlab{}.
\newblock \showarticletitle{On the number of inference regions of deep feed
  forward networks with piece-wise linear activations}. In
  \bibinfo{booktitle}{\emph{International Conference on Learning
  Representations (ICLR)}}.
\newblock


\bibitem[P{\'e}bay et~al\mbox{.}(2007)]%
        {pebay2007new}
\bibfield{author}{\bibinfo{person}{Philippe~P P{\'e}bay},
  \bibinfo{person}{David Thompson}, \bibinfo{person}{Jason Shepherd},
  \bibinfo{person}{Patrick Knupp}, \bibinfo{person}{Curtis Lisle},
  \bibinfo{person}{Vincent~A Magnotta}, {and} \bibinfo{person}{Nicole~M
  Grosland}.} \bibinfo{year}{2007}\natexlab{}.
\newblock \showarticletitle{New applications of the verdict library for
  standardized mesh verification pre, post, and end-to-end processing}. In
  \bibinfo{booktitle}{\emph{International Meshing Roundtable}}.
  \bibinfo{pages}{535--552}.
\newblock


\bibitem[Raghu et~al\mbox{.}(2017)]%
        {raghu2017expressive}
\bibfield{author}{\bibinfo{person}{Maithra Raghu}, \bibinfo{person}{Ben Poole},
  \bibinfo{person}{Jon Kleinberg}, \bibinfo{person}{Surya Ganguli}, {and}
  \bibinfo{person}{Jascha Sohl-Dickstein}.} \bibinfo{year}{2017}\natexlab{}.
\newblock \showarticletitle{On the expressive power of deep neural networks}.
  In \bibinfo{booktitle}{\emph{International Conference on Machine Learning
  (ICML)}}. \bibinfo{pages}{2847--2854}.
\newblock


\bibitem[Ren et~al\mbox{.}(2025)]%
        {ren2025mcgrids}
\bibfield{author}{\bibinfo{person}{Daxuan Ren}, \bibinfo{person}{Hezi Shi},
  \bibinfo{person}{Jianmin Zheng}, {and} \bibinfo{person}{Jianfei Cai}.}
  \bibinfo{year}{2025}\natexlab{}.
\newblock \showarticletitle{{McGrids}: Monte Carlo-Driven Adaptive Grids for
  Iso-Surface Extraction}. In \bibinfo{booktitle}{\emph{European Conference on
  Computer Vision (ECCV)}}. Springer, \bibinfo{pages}{127--144}.
\newblock


\bibitem[Rump and Kashiwagi(2015)]%
        {rump2015implementation}
\bibfield{author}{\bibinfo{person}{Siegfried~M Rump} {and}
  \bibinfo{person}{Masahide Kashiwagi}.} \bibinfo{year}{2015}\natexlab{}.
\newblock \showarticletitle{Implementation and improvements of affine
  arithmetic}.
\newblock \bibinfo{journal}{\emph{Nonlinear Theory and Its Applications,
  IEICE}} \bibinfo{volume}{6}, \bibinfo{number}{3} (\bibinfo{year}{2015}),
  \bibinfo{pages}{341--359}.
\newblock


\bibitem[Rupert(1995)]%
        {rupert1995delaunay}
\bibfield{author}{\bibinfo{person}{Jim Rupert}.}
  \bibinfo{year}{1995}\natexlab{}.
\newblock \showarticletitle{A Delaunay refinement algorithm for quality 2D-mesh
  generation}.
\newblock \bibinfo{journal}{\emph{Journal of Algorithms}} \bibinfo{volume}{18},
  \bibinfo{number}{3} (\bibinfo{year}{1995}), \bibinfo{pages}{548--585}.
\newblock


\bibitem[Sell{\'a}n et~al\mbox{.}(2023)]%
        {sellan2023reach}
\bibfield{author}{\bibinfo{person}{Silvia Sell{\'a}n},
  \bibinfo{person}{Christopher Batty}, {and} \bibinfo{person}{Oded Stein}.}
  \bibinfo{year}{2023}\natexlab{}.
\newblock \showarticletitle{Reach For the Spheres: Tangency-aware surface
  reconstruction of SDFs}. In \bibinfo{booktitle}{\emph{SIGGRAPH Asia}}.
\newblock


\bibitem[Sell\'{a}n et~al\mbox{.}(2024)]%
        {Sellan2024RFTA}
\bibfield{author}{\bibinfo{person}{Silvia Sell\'{a}n},
  \bibinfo{person}{Yingying Ren}, \bibinfo{person}{Christopher Batty}, {and}
  \bibinfo{person}{Oded Stein}.} \bibinfo{year}{2024}\natexlab{}.
\newblock \showarticletitle{Reach For the Arcs: Reconstructing Surfaces from
  SDFs via Tangent Points}. In \bibinfo{booktitle}{\emph{SIGGRAPH}}. Article
  \bibinfo{articleno}{25}.
\newblock


\bibitem[Serra et~al\mbox{.}(2018)]%
        {serra2018bounding}
\bibfield{author}{\bibinfo{person}{Thiago Serra}, \bibinfo{person}{Christian
  Tjandraatmadja}, {and} \bibinfo{person}{Srikumar Ramalingam}.}
  \bibinfo{year}{2018}\natexlab{}.
\newblock \showarticletitle{Bounding and counting linear regions of deep neural
  networks}. In \bibinfo{booktitle}{\emph{International Conference on Machine
  Learning (ICML)}}. \bibinfo{pages}{4558--4566}.
\newblock


\bibitem[Seyb et~al\mbox{.}(2019)]%
        {seyb2019non}
\bibfield{author}{\bibinfo{person}{Dario Seyb}, \bibinfo{person}{Alec
  Jacobson}, \bibinfo{person}{Derek Nowrouzezahrai}, {and}
  \bibinfo{person}{Wojciech Jarosz}.} \bibinfo{year}{2019}\natexlab{}.
\newblock \showarticletitle{Non-linear sphere tracing for rendering deformed
  signed distance fields}.
\newblock \bibinfo{journal}{\emph{ACM Transactions on Graphics (TOG)}}
  \bibinfo{volume}{38}, \bibinfo{number}{6} (\bibinfo{year}{2019}).
\newblock


\bibitem[Sharp and Jacobson(2022)]%
        {sharp2022spelunking}
\bibfield{author}{\bibinfo{person}{Nicholas Sharp} {and} \bibinfo{person}{Alec
  Jacobson}.} \bibinfo{year}{2022}\natexlab{}.
\newblock \showarticletitle{Spelunking the deep: Guaranteed queries on general
  neural implicit surfaces via range analysis}.
\newblock \bibinfo{journal}{\emph{ACM Transactions on Graphics (TOG)}}
  \bibinfo{volume}{41}, \bibinfo{number}{4} (\bibinfo{year}{2022}),
  \bibinfo{pages}{1--16}.
\newblock


\bibitem[Shen et~al\mbox{.}(2023)]%
        {shen2023flexicubes}
\bibfield{author}{\bibinfo{person}{Tianchang Shen}, \bibinfo{person}{Jacob
  Munkberg}, \bibinfo{person}{Jon Hasselgren}, \bibinfo{person}{Kangxue Yin},
  \bibinfo{person}{Zian Wang}, \bibinfo{person}{Wenzheng Chen},
  \bibinfo{person}{Zan Gojcic}, \bibinfo{person}{Sanja Fidler},
  \bibinfo{person}{Nicholas Sharp}, {and} \bibinfo{person}{Jun Gao}.}
  \bibinfo{year}{2023}\natexlab{}.
\newblock \showarticletitle{Flexible Isosurface Extraction for Gradient-Based
  Mesh Optimization}.
\newblock \bibinfo{journal}{\emph{ACM Transactions on Graphics (TOG)}}
  \bibinfo{volume}{42}, \bibinfo{number}{4} (\bibinfo{year}{2023}).
\newblock


\bibitem[Shewchuk(1999)]%
        {shewchuk1999lecture}
\bibfield{author}{\bibinfo{person}{Jonathan~Richard Shewchuk}.}
  \bibinfo{year}{1999}\natexlab{}.
\newblock \showarticletitle{Lecture notes on Delaunay mesh generation}.
\newblock  (\bibinfo{year}{1999}).
\newblock


\bibitem[Sitzmann et~al\mbox{.}(2020)]%
        {sitzmann2019siren}
\bibfield{author}{\bibinfo{person}{Vincent Sitzmann},
  \bibinfo{person}{Julien~N.P. Martel}, \bibinfo{person}{Alexander~W. Bergman},
  \bibinfo{person}{David~B. Lindell}, {and} \bibinfo{person}{Gordon
  Wetzstein}.} \bibinfo{year}{2020}\natexlab{}.
\newblock \showarticletitle{Implicit Neural Representations with Periodic
  Activation Functions}. In \bibinfo{booktitle}{\emph{Advances in Neural
  Information Processing Systems (NeurIPS)}}.
\newblock


\bibitem[Sitzmann et~al\mbox{.}(2019)]%
        {sitzmann2019scene}
\bibfield{author}{\bibinfo{person}{Vincent Sitzmann}, \bibinfo{person}{Michael
  Zollh{\"o}fer}, {and} \bibinfo{person}{Gordon Wetzstein}.}
  \bibinfo{year}{2019}\natexlab{}.
\newblock \showarticletitle{Scene representation networks: Continuous
  3d-structure-aware neural scene representations}.
\newblock \bibinfo{journal}{\emph{Advances in Neural Information Processing
  Systems (NeurIPS)}}  \bibinfo{volume}{32} (\bibinfo{year}{2019}).
\newblock


\bibitem[Sorgente et~al\mbox{.}(2023)]%
        {sorgente2023survey}
\bibfield{author}{\bibinfo{person}{Tommaso Sorgente}, \bibinfo{person}{Silvia
  Biasotti}, \bibinfo{person}{Gianmarco Manzini}, {and}
  \bibinfo{person}{Michela Spagnuolo}.} \bibinfo{year}{2023}\natexlab{}.
\newblock \showarticletitle{A survey of indicators for mesh quality
  assessment}. In \bibinfo{booktitle}{\emph{Computer Graphics Forum}},
  Vol.~\bibinfo{volume}{42}. \bibinfo{pages}{461--483}.
\newblock


\bibitem[Stander and Hart(1997)]%
        {stander1997guaranteeing}
\bibfield{author}{\bibinfo{person}{Barton~T Stander} {and}
  \bibinfo{person}{John~C Hart}.} \bibinfo{year}{1997}\natexlab{}.
\newblock \showarticletitle{Guaranteeing the topology of an implicit surface
  polygonization for interactive modeling}. In
  \bibinfo{booktitle}{\emph{Conference on Computer Graphics and Interactive
  Techniques}}. \bibinfo{pages}{279--286}.
\newblock


\bibitem[Stimpson et~al\mbox{.}(2007)]%
        {stimpson2007verdict}
\bibfield{author}{\bibinfo{person}{CJ Stimpson}, \bibinfo{person}{CD Ernst},
  \bibinfo{person}{David~C Thompson}, \bibinfo{person}{Patrick~Michael Knupp},
  {and} \bibinfo{person}{Philippe~Pierre P{\'e}bay}.}
  \bibinfo{year}{2007}\natexlab{}.
\newblock \bibinfo{booktitle}{\emph{The verdict geometric quality library}}.
\newblock Number 1751. \bibinfo{publisher}{Sandia National Laboratories}.
\newblock


\bibitem[Sutherland and Hodgman(1974)]%
        {sutherland1974reentrant}
\bibfield{author}{\bibinfo{person}{Ivan~E Sutherland} {and}
  \bibinfo{person}{Gary~W Hodgman}.} \bibinfo{year}{1974}\natexlab{}.
\newblock \showarticletitle{Reentrant polygon clipping}.
\newblock \bibinfo{journal}{\emph{Commun. ACM}} \bibinfo{volume}{17},
  \bibinfo{number}{1} (\bibinfo{year}{1974}), \bibinfo{pages}{32--42}.
\newblock


\bibitem[Tancik et~al\mbox{.}(2020)]%
        {tancik2020fourier}
\bibfield{author}{\bibinfo{person}{Matthew Tancik}, \bibinfo{person}{Pratul
  Srinivasan}, \bibinfo{person}{Ben Mildenhall}, \bibinfo{person}{Sara
  Fridovich-Keil}, \bibinfo{person}{Nithin Raghavan}, \bibinfo{person}{Utkarsh
  Singhal}, \bibinfo{person}{Ravi Ramamoorthi}, \bibinfo{person}{Jonathan
  Barron}, {and} \bibinfo{person}{Ren Ng}.} \bibinfo{year}{2020}\natexlab{}.
\newblock \showarticletitle{Fourier features let networks learn high frequency
  functions in low dimensional domains}.
\newblock \bibinfo{journal}{\emph{Advances in Neural Information Processing
  Systems (NeurIPS)}}  \bibinfo{volume}{33} (\bibinfo{year}{2020}),
  \bibinfo{pages}{7537--7547}.
\newblock


\bibitem[Vallin et~al\mbox{.}(2023)]%
        {vallin2023geometric}
\bibfield{author}{\bibinfo{person}{Jonatan Vallin}, \bibinfo{person}{Karl
  Larsson}, {and} \bibinfo{person}{Mats~G Larson}.}
  \bibinfo{year}{2023}\natexlab{}.
\newblock \showarticletitle{The geometric structure of fully-connected
  relu-layers}.
\newblock \bibinfo{journal}{\emph{arXiv preprint arXiv:2310.03482}}
  (\bibinfo{year}{2023}).
\newblock


\bibitem[Van~Overveld and Wyvill(2004)]%
        {van2004shrinkwrap}
\bibfield{author}{\bibinfo{person}{Kees Van~Overveld} {and}
  \bibinfo{person}{Brian Wyvill}.} \bibinfo{year}{2004}\natexlab{}.
\newblock \showarticletitle{Shrinkwrap: An efficient adaptive algorithm for
  triangulating an iso-surface}.
\newblock \bibinfo{journal}{\emph{The Visual Computer}}  \bibinfo{volume}{20}
  (\bibinfo{year}{2004}), \bibinfo{pages}{362--379}.
\newblock


\bibitem[Vincent and Schwager(2021)]%
        {vincent2021RPM}
\bibfield{author}{\bibinfo{person}{Joseph~A. Vincent} {and}
  \bibinfo{person}{Mac Schwager}.} \bibinfo{year}{2021}\natexlab{}.
\newblock \showarticletitle{Reachable Polyhedral Marching (RPM): A Safety
  Verification Algorithm for Robotic Systems with Deep Neural Network
  Components}. In \bibinfo{booktitle}{\emph{International Conference on
  Robotics and Automation (ICRA)}}. \bibinfo{pages}{9029--9035}.
\newblock


\bibitem[Wang et~al\mbox{.}(2021)]%
        {wang2021neus}
\bibfield{author}{\bibinfo{person}{Peng Wang}, \bibinfo{person}{Lingjie Liu},
  \bibinfo{person}{Yuan Liu}, \bibinfo{person}{Christian Theobalt},
  \bibinfo{person}{Taku Komura}, {and} \bibinfo{person}{Wenping Wang}.}
  \bibinfo{year}{2021}\natexlab{}.
\newblock \showarticletitle{NeuS: Learning Neural Implicit Surfaces by Volume
  Rendering for Multi-view Reconstruction}.
\newblock \bibinfo{journal}{\emph{Advances in Neural Information Processing
  Systems (NeurIPS)}} (\bibinfo{year}{2021}).
\newblock


\bibitem[Wang(2022)]%
        {wang2022estimation}
\bibfield{author}{\bibinfo{person}{Yuan Wang}.}
  \bibinfo{year}{2022}\natexlab{}.
\newblock \showarticletitle{Estimation and Comparison of Linear Regions for
  ReLU Networks}. In \bibinfo{booktitle}{\emph{International Joint Conferences
  on Artificial Intelligence (IJCAI)}}. \bibinfo{pages}{3544--3550}.
\newblock


\bibitem[Watt(1999)]%
        {watt19993d}
\bibfield{author}{\bibinfo{person}{Alan~H Watt}.}
  \bibinfo{year}{1999}\natexlab{}.
\newblock \bibinfo{booktitle}{\emph{3{D} Computer Graphics}}.
\newblock \bibinfo{publisher}{Addison-Wesley Longman Publishing Co., Inc.}
\newblock


\bibitem[Wilhelms and Van~Gelder(1992)]%
        {wilhelms1992octrees}
\bibfield{author}{\bibinfo{person}{Jane Wilhelms} {and} \bibinfo{person}{Allen
  Van~Gelder}.} \bibinfo{year}{1992}\natexlab{}.
\newblock \showarticletitle{Octrees for faster isosurface generation}.
\newblock \bibinfo{journal}{\emph{ACM Transactions on Graphics (TOG)}}
  \bibinfo{volume}{11}, \bibinfo{number}{3} (\bibinfo{year}{1992}),
  \bibinfo{pages}{201--227}.
\newblock


\bibitem[Wyvill et~al\mbox{.}(1986)]%
        {wyvill1986data}
\bibfield{author}{\bibinfo{person}{Geoff Wyvill}, \bibinfo{person}{Craig
  McPheeters}, {and} \bibinfo{person}{Brian Wyvill}.}
  \bibinfo{year}{1986}\natexlab{}.
\newblock \showarticletitle{Data structure for soft objects}.
\newblock \bibinfo{journal}{\emph{The Visual Computer}}  \bibinfo{volume}{2}
  (\bibinfo{year}{1986}), \bibinfo{pages}{227--234}.
\newblock


\bibitem[Xie et~al\mbox{.}(2022)]%
        {xie2022neural}
\bibfield{author}{\bibinfo{person}{Yiheng Xie}, \bibinfo{person}{Towaki
  Takikawa}, \bibinfo{person}{Shunsuke Saito}, \bibinfo{person}{Or Litany},
  \bibinfo{person}{Shiqin Yan}, \bibinfo{person}{Numair Khan},
  \bibinfo{person}{Federico Tombari}, \bibinfo{person}{James Tompkin},
  \bibinfo{person}{Vincent Sitzmann}, {and} \bibinfo{person}{Srinath Sridhar}.}
  \bibinfo{year}{2022}\natexlab{}.
\newblock \showarticletitle{Neural fields in visual computing and beyond}. In
  \bibinfo{booktitle}{\emph{Computer Graphics Forum}},
  Vol.~\bibinfo{volume}{41}. \bibinfo{pages}{641--676}.
\newblock


\bibitem[Zhou and Jacobson(2016)]%
        {Thingi10K}
\bibfield{author}{\bibinfo{person}{Qingnan Zhou} {and} \bibinfo{person}{Alec
  Jacobson}.} \bibinfo{year}{2016}\natexlab{}.
\newblock \showarticletitle{Thingi10K: A Dataset of 10,000 3D-Printing Models}.
\newblock \bibinfo{journal}{\emph{arXiv preprint arXiv:1605.04797}}
  (\bibinfo{year}{2016}).
\newblock


\bibitem[Zhou et~al\mbox{.}(2018)]%
        {Zhou2018open3d}
\bibfield{author}{\bibinfo{person}{Qian-Yi Zhou}, \bibinfo{person}{Jaesik
  Park}, {and} \bibinfo{person}{Vladlen Koltun}.}
  \bibinfo{year}{2018}\natexlab{}.
\newblock \showarticletitle{{Open3D}: {A} Modern Library for {3D} Data
  Processing}.
\newblock \bibinfo{journal}{\emph{arXiv:1801.09847}} (\bibinfo{year}{2018}).
\newblock


\end{thebibliography}

\end{document}